\documentstyle[12pt]{article}
\newcommand {\Obs} {{\cal O}}
\newlength{\digitwidth} 

\def\sqr#1#2{{\vcenter{\hrule height.#2pt
      \hbox{\vrule width.#2pt height#1pt \kern#1pt
        \vrule width.#2pt}
      \hrule height.#2pt}}}

\def\abstracts#1#2#3{{
        \centering{\begin{minipage}{4.62in}\baselineskip=13pt
        \small
        \centerline{\bf Abstract}
        \vspace*{0.2cm}                
        \parindent=0pt #1\par
        \parindent=18pt #2\par
        \parindent=15pt #3
        \end{minipage} }\par}}
\def\keywords#1#2#3{{
        \centering{\begin{minipage}{4.62in}\baselineskip=13pt
        \small
        \vspace*{0.2cm}                
        \parindent=0pt #1\par
        \parindent=18pt #2\par
        \parindent=15pt #3
        \end{minipage} }\par}}
\renewcommand{\thefootnote}{\fnsymbol{footnote}}
\begin{document}
\vspace*{-3cm}
\hfill \parbox{4.2cm}{ FUB-HEP 15/93\\
                August 1993\\
                revised March 1994 \\ 
               }\\
\vspace*{1.3cm}
\centerline{\LARGE \bf Application of the multicanonical} \\[0.2cm]
\centerline{\LARGE \bf multigrid Monte Carlo method to} \\[0.2cm]
\centerline{\LARGE \bf the two-dimensional $\phi^4$-model:}\\[0.2cm]
\centerline{\LARGE \bf autocorrelations and interface 
                       tension\footnotemark}\\[0.2cm]
\footnotetext{\noindent Work supported in part by Deutsche 
Forschungsgemeinschaft under grant Kl256.}
\addtocounter{footnote}{-1}
\renewcommand{\thefootnote}{\arabic{footnote}}
\vspace*{0.1cm}
\centerline{\large {\em Wolfhard Janke\/}$^{1}$ 
               and {\em Tilman Sauer\/}$^2$}\\[0.4cm] 
\centerline{\large    $^1$ {\small Institut f\"ur Physik,
                      Johannes Gutenberg-Universit\"at Mainz}}
\centerline{    {\small 55099 Mainz, Germany }}\\[0.15cm]
\centerline{\large    $^2$ {\small Institut f\"{u}r Theoretische Physik,
                      Freie Universit\"{a}t Berlin}}
\centerline{    {\small 14195 Berlin, Germany}}\\[2.50cm]
\vspace*{0.3cm}
\abstracts{}{
We discuss the recently proposed multicanonical multigrid
Monte Carlo method and apply it to the scalar
$\phi^4$-model on a square lattice.
To investigate the performance of the new algorithm at the 
field-driven first-order phase transitions between the two 
ordered phases we carefully analyze the autocorrelations of the 
Monte Carlo process. 
Compared with standard multicanonical simulations a real-time 
improvement of about one order of magnitude is established.
The interface tension between the two ordered phases is extracted
from high-statistics histograms of the magnetization
applying histogram reweighting techniques.
}{}
\vspace*{0.5cm}
\keywords{}{
}{}
  \thispagestyle{empty}
      \newpage
%
       \section{Introduction}
%
First-order phase transitions play an important role in many
fields of physics \cite{gunton,juelich92}. 
Examples range from the well-known process of crystal melting through 
the deconfining transition in hot quark-gluon matter to various steps 
in the evolution of the early universe. 
It is therefore gratifying that recently high precision Monte Carlo 
studies of systems undergoing a first-order phase transition have 
become feasible by showing that the problem of the supercritical 
slowing down, governed by exponentially diverging autocorrelation times
\begin{equation}
   \tau \propto \exp\{2\sigma L^{d-1}\},
  \label{eq:scsd}
\end{equation}
may be eliminated by means of the so-called multicanonical algorithm
\cite{muca1}.
Here $\sigma$ is the interface tension between the coexisting phases, 
$L$ the linear size of a $d$-dimensional cubic system, and the factor
$2$ accounts for the usually employed periodic boundary conditions.

While the multicanonical
algorithm does beat the exponential slowing down the remaining
autocorrelation times typically still diverge with
some power $\alpha \approx 1 \dots 1.5$ of the lattice volume $V=L^d$
\cite{muca1,jbk92,multimagnetical,muca2a,muca2b,muca3,muca4},
\begin{equation}
   \tau \propto V^\alpha,
  \label{eq:csd}
\end{equation}
and may consequently 
still be severe. It is therefore worthwhile to look for further 
improvements of the Monte Carlo scheme. 
Critical slowing down with a power-law divergence of the autocorrelation
time is a notorious problem for simulating systems at a second-order
phase transition. Fortunately also here substantial progress was made
in the past few years by designing modified Monte Carlo update schemes 
which reduce or even eliminate the critical slowing down,
see Ref.\cite{review} for reviews. 
Among these sophisticated Monte Carlo schemes multigrid techniques 
\cite{mgmc1,mgmc2,mgmc3,mgmc4,js93a,js93b} 
have been shown both to perform quite successfully and to be rather 
generally applicable.
In a recent paper we have therefore proposed \cite{js93b,js93c} to combine the 
multicanonical approach with multigrid techniques and have presented
preliminary investigations \cite{js93c} which show that autocorrelation
times in the 
multicanonical simulation may further be reduced by this combination.
The purpose of this paper is to present a detailed study of the
performance of the multicanonical multigrid method applied to
a scalar two-dimensional $\phi^4$-theory on the lattice.

For Potts models it was recently proposed along other lines 
to combine a multicanonical demon algorithm with cluster update methods
in a hybrid-like fashion \cite{rummukainen93}.  Another interesting idea
is to enhance the performance of Monte Carlo simulations of systems at 
a first-order phase transition by treating the parameter which controls 
the strength of the transition as a dynamical variable \cite{kw93}.

The outline of the paper is as follows.
After defining the model and discussing its basic features in section 2 
we will briefly review the characteristic properties of multicanonical 
reweighting and of multigrid update techniques in section 3, 
and show how the two Monte Carlo schemes may be combined.
We further discuss the error estimates for canonical observables 
computed from multicanonical simulations and introduce an associated 
effective autocorrelation time which allows for a direct comparison 
with canonical simulations. 
Details of the calculation which in principle is straightforward but 
nonetheless requires some care are presented in Appendix A.
In section 4 we focus on the autocorrelation times achieved by the 
different update algorithms. After presenting some data for the 
canonical case for later comparison we first analyze autocorrelations 
in the pure multicanonical distribution. We then discuss how the 
effective autocorrelation time emerges from these data.
Since the multicanonical multigrid method allows to simulate the model 
with some accuracy, we compute in section 5 the interface tension using 
histogram reweighting techniques. 
In section 6 we conclude with some remarks on the applicability of the 
method and on future perspectives. 

%
       \section{The model and observables}
%

As a test case to study the performance of the recently proposed 
multicanonical multigrid algorithm \cite{js93b,js93c} we have taken the 
two-dimensional scalar $\phi^4$-model on a square $L \times L$-lattice
with periodic boundary conditions.
This model has been studied recently in a number of numerical
investigations and has repeatedly been used as a testing ground for
advanced numerical simulation techniques 
\cite{mgmc1,mhb86,bt89,tc90,mf92}. 
Previous investigations have focussed on properties of the model at 
criticality \cite{tc90}, in particular on the question of
finite-size scaling and the universality with the Ising 
model \cite{mhb86,mf92}. 
In this paper we present data from simulations performed in the broken
symmetry phase, 
that is, {\em strictly below} the critical temperature at zero
magnetic field (along the first-order transition curve).
We focus 
on the autocorrelations of the Monte Carlo process and on 
the interface tension between the two ordered phases of positive and 
negative magnetization.

The model is defined by the partition function
\begin{equation}
      {\cal Z}(\mu^2, g) = 
        \prod_{i=1}^{V} \left[ \int_{-\infty}^{+\infty} d\phi_i\right] 
          \exp\{-{\cal H}(\{\phi_i\})\} ,
 \label{eq:model1}
\end{equation}
where
\begin{equation}
          {\cal H}(\{\phi_i\}) = 
               \sum_{i=1}^V[\frac{1}{2}(\vec{\nabla}\phi_i)^2
               - \frac{\mu^2}{2}\phi_i^2 + g\phi_i^4].
 \label{eq:model2}
\end{equation}
Here $(\vec{\nabla}\phi_i)^2 = \sum_{\mu=1}^d (\phi_{i_{\mu}}-\phi_i)^2$ 
is the squared lattice gradient where $i_{\mu}$ denotes the nearest 
neighbor index in the lattice direction $\mu$, and $V=L^d$ is the
volume of a $d$-dimensional cubic system.
The constants $\mu^2,g > 0$ are parameters to be specified later, and
the energy is measured in temperature units, i.e., we have set 
$\beta = 1/k_BT = 1$.
Observables for the model are denoted by  
\begin{equation}
      M_k \equiv \sum_{i=1}^{V} \phi_i^k, \quad k=1,2,\dots, 
 \label{eq:notation2}
\end{equation}
and the corresponding densities are denoted by $m_i \equiv M_i / V$.
The kinetic term will be denoted by 
\begin{equation}
      K_0 \equiv \sum_{i=1}^{V} (\vec{\nabla}\phi_i)^2. 
 \label{eq:notation1}
\end{equation}
Other quantities can be defined as functions of these
observables, thus the energy $E$ is given by
\begin{eqnarray}
    E & = & \frac{1}{2}K_0 - \frac{\mu^2}{2}M_2 + gM_4,
\end{eqnarray}
and the specific heat $C$ and the (finite lattice)
susceptibility $\chi$ can be obtained from 
\begin{eqnarray}
    C & = & ( \langle E^2 \rangle - \langle E \rangle^2 ) /V, \\
    \chi & = & ( \langle M_1^2 \rangle - \langle |M_1| \rangle^2 ) /V. 
\end{eqnarray}

In $d=2$ dimensions the model (\ref{eq:model1},\ref{eq:model2}) displays 
a line of second-order phase transitions in the $(\mu^2,g)$-plane which 
was numerically determined in Ref.\cite{tc90}.
In the thermodynamic limit the system shows spontaneous symmetry
breaking for all $\mu^2 > \mu_c^2(g)$ with a nonvanishing expectation 
value of the average magnetization $\langle M_1 \rangle$. 
Adding a term $-h\phi_i$ to the 
energy (\ref{eq:model2}) the system changes discontinuously from a state 
of positive magnetization $\langle M_1 \rangle > 0$ to a state of 
negative magnetization $\langle M_1 \rangle < 0$ if we tune the magnetic
field $h$ from positive to negative values.
For vanishing magnetic field $h=0$, the system consequently is at a 
first-order phase transition. If the linear length is finite, 
$L<\infty$, the system then is flipping back and forth between states 
of positive and negative magnetization which renders 
$\langle M_1 \rangle = 0$ also for $\mu^2 > \mu_c^2(g)$. At a
first-order phase
transition point finite systems can also exist in a mixed phase 
configuration which is characterized by regions of the pure phases
separated by interfaces. For topological reasons
there are necessarily an even number of interfaces of length $L$ 
for periodic boundary conditions which we have always used.
For energetic reasons configurations of more than two interfaces
almost never occur.
A typical mixed phase configuration is shown in Fig.1.
Due to the additional free energy of the interfaces, configurations
with small total magnetization are suppressed by a factor 
$\exp(-2\sigma L)$ where $\sigma$ is the interface tension.
For this reason the probability distribution of the magnetization 
$P(m_1)$ shows two distinct peaks separated by a region of
strongly suppressed mixed phase configurations.

Following Binder \cite{binder81} one can 
extract the interface tension $\sigma$ by determining the ratio
of the maxima $P^{\rm max}$ to the minimum $P^{\rm min}$ of the 
distribution $P(m_1)$. 
The interface tension is then given by
\begin{equation}
    \sigma = \lim_{L\to\infty}\sigma_L,
 \label{extrapolation}
\end{equation}
with
\begin{equation}
    e^{-2\sigma_L L} = \frac{P^{\rm min}}{P^{\rm max}}.
\end{equation}
This expression can easily be evaluated provided that the statistical
uncertainties of both $P^{\rm max}$ and $P^{\rm min}$ are small,
which can be achieved by performing simulations according to the
multicanonical distribution.

Since in this paper we study the $\phi^4$-model as a testing ground 
for the performance of Monte Carlo algorithms at first-order phase 
transitions the primary observable of interest will be the average 
magnetization $m_1$ and its autocorrelations in the Monte Carlo
process. Although for the partition function 
(\ref{eq:model1},\ref{eq:model2}) this observable vanishes trivially 
in finite systems
for reasons of symmetry we emphasize that this symmetry is completely
artificial and would at once be broken, e.g., by adding some
term of odd power in ${\phi}$ to the energy in (\ref{eq:model2}).

Autocorrelations provide a measure for the dynamics of 
different Monte Carlo schemes, see Ref.\cite{ms88} for a review.
In general, if $\Obs_i$ denotes the $i$-th measurement of an observable
$\Obs$ in the Monte Carlo process the autocorrelation function $A(j)$
is defined by 
\begin{equation}
      A(j) = \frac{ \langle \Obs_i\Obs_{i+j} \rangle 
                     - \langle \Obs_i \rangle^2 }
                  { \langle \Obs_i^2 \rangle
                     - \langle \Obs_i \rangle^2 },
 \label{eq:A(j)}
\end{equation}
and from $A(j)$ the integrated autocorrelation time $\tau^{\rm int}$ is obtained 
in the large $k$ limit of
\begin{equation}
      \tau(k) = \frac{1}{2} + \sum_{j=1}^k A(j).
  \label{eq:Bdef} 
\end{equation}
Since for large $j$ the autocorrelation function $A(j)$ decays like an 
exponential
\begin{equation}
      A(j) \buildrel j \rightarrow \infty \over \longrightarrow
            a \exp(-j/\tau^{\rm exp}),
\end{equation}
where $\tau^{\rm exp}$ denotes the exponential autocorrelation time and $a$ is
some constant, $\tau(k)$ behaves like 
\begin{eqnarray}
      \tau(k) & = & \tau^{\rm int} -
                     a \sum_{j=k+1}^{\infty} \exp(-j/\tau^{\rm exp}) 
        \\ 
           & = & \tau^{\rm int} - a \frac{\exp\{-1/\tau^{\rm exp}\}}
               {1-\exp\{ -1/\tau^{\rm exp} \}}
                \exp\{-k/\tau^{\rm exp}\},
           \label{eq:Bfit} 
\end{eqnarray}
where $\tau^{\rm int} \equiv \tau(\infty)$.
The latter expression may be used for a numerical determination
of the exponential and integrated autocorrelation times.
Since, in general, all these quantities depend on the observable under
consideration we will indicate the relevant observable by an additional
subscript unless it is clear from the context which obserable is meant.

%
       \section{Multicanonical Monte Carlo simulations using
                multigrid techniques}

\subsection[]{Multicanonical simulations}

The basic idea of the multicanonical approach 
\cite{muca1,muca3,muca4} is to simulate an
auxiliary distribution in which the mixed phase configurations
have the same weight as the pure phases and canonical expectation
values are recovered by reweighting.
Hence the multicanonical approach is not itself a Monte Carlo update 
algorithm but a general reweighting prescription which allows to
simulate distributions which are numerically easier to handle.

While similar ideas have been known in 
the literature under the name of umbrella sampling already for a long
time \cite{umbrella} the practical relevance of multicanonical
reweighting techniques for simulations of first-order phase transitions 
\cite{muca1,jbk92,multimagnetical,muca2a,muca2b,muca3,muca4}
was realized only recently
\cite{muca1}. The multicanonical approach may be formulated in 
a rather general way for a variety of applications \cite{muca3} 
but in the context of first-order phase transitions and quantum
mechanical tunneling problems \cite{js93b,js93c}
it may simply be regarded as being basically a reweighting 
technique \cite{muca4}.

Let $m = m(\{\phi_i\})$  be an observable whose probability distribution 
in the canonical ensemble displays two strong separated peaks.
In a field driven transition, as in our case, $m(\{\phi_i\})$ is the 
magnetization $m_1$ and the situation has also been
referred to as multimagnetical simulation \cite{multimagnetical}. 
For a temperature
driven transition the relevant observable would be the energy of the 
system. In the multicanonical reweighting approach 
we now rewrite the partition function by introducing some
function $f$ as 
\begin{equation}
  {\cal Z} = \prod_{i=1}^V \int d \phi_i e^{-{\cal H}(\{\phi_i\})-f(m)}e^{f(m)},
\end{equation}
and adjust the reweighting factor $\exp(-f(m))$ in such a way that the 
resulting histogram of $m$ sampled according to the 
{\em multicanonical} probability distribution
\begin{eqnarray}
    p^{\rm muca}(\{\phi_i\}) &\propto& \exp[-{\cal H}(\{\phi_i\})-f(m)]
                    \nonumber \\
                             &\equiv&  \exp[-{\cal H}^{\rm muca}(\{\phi_i\})]
 \label{eq:mucaenergy}
\end{eqnarray}
is approximately flat. 
Here ${\cal H}^{\rm muca}$ is the central object of a multicanonical 
simulation, and plays the same role in it as ${\cal H}$ does in a 
canonical simulation.
Canonical observables $\langle {\cal O} \rangle_{\rm can}$ can be 
recovered according to 
\begin{equation}
    \langle {\cal O} \rangle_{\rm can} = 
        \frac{\langle {\cal O}w \rangle}{\langle w \rangle},
 \label{eq:mucareweight}
\end{equation}
where $\langle \dots \rangle$ without subscript denote expectation 
values in the
multicanonical distribution and $w \equiv \exp(f(m))$ is the inverse 
reweighting factor. 

The multicanonical probability distribution $p^{\rm muca}$ may be updated
using any 
legitimate Monte Carlo algorithm. The simplest choice is a local 
Metropolis update 
where we consider as usual local moves
$\phi_{i_0} \rightarrow \phi_{i_0} + \Delta\phi_{i_0}$ at some site ${i_0}$ 
and compute the energy difference $\Delta E^{\rm muca}$ according to 
(\ref{eq:mucaenergy}), i.e., the
decision on whether a Metropolis move will be accepted or not is now 
to be based on the energy difference
\begin{equation}
     \Delta E^{\rm muca} = \Delta E + f(m + \Delta m) - f(m),
\end{equation}  
where 
$\Delta E={\cal H}(\phi_1,\phi_2,\dots,\phi_{i_0}+\Delta\phi_{i_0},\dots,
\phi_V) - {\cal H}(\{\phi_i\})$
is the canonical energy difference and $\Delta m = m(\phi_1,\phi_2,
\dots, \phi_{i_0}+\Delta \phi_{i_0}, \dots, \phi_V)-m(\{\phi_i\})$
is the corresponding difference in the observable $m$.
If the canonical probability distribution is reweighted in the
magnetization, $m(\{\phi_i\}) = m_1 = \sum_{i=1}^V \phi_i/V$,
this difference is simply given by $\Delta m = \Delta \phi_{i_0}/V$.
For a temperature-driven transition we have $m = {\cal H}/V$ and 
$\Delta m = \Delta E/V$.

In practical simulations $f$ may be recorded in form of a simple stair 
case function which does not introduce any numerical inaccuracy
since this factor cancels out in all canonical expectation values.
It is also worth mentioning that since $m$ depends on
all values of $\phi_i$, the resulting multicanonical energy is
essentially {\em non\/}local. 

As we will discuss in Section 3.3,
the multicanonical probability distribution $p^{\rm muca}$ may also be
updated by a multigrid Monte Carlo method.

\subsection[]{Multigrid techniques}

The basic idea of multigrid Monte Carlo techniques 
\cite{mgmc1,mgmc2,mgmc3,mgmc4,js93a,js93b,hac85,cor88}
is to systematically perform updates on different length scales of the 
system. In the corresponding {\em uni}\/grid viewpoint, which always 
looks at the effects on the original fine grained lattice, this is done 
by moving blocks of $1$, $2^d$, $4^d$, $8^d$, $\dots$, $2^{nd}=V$ 
adjacent variables at a time. 
In the multigrid formulation these collective update moves are
implemented by introducing auxiliary fields on coarse-grained
lattices. Specifically one introduces a sequence of coarsened grids
$\Xi^{(k)}, k = n-1, \dots, 0$ of size $2^{kd}$. In the simplest 
piecewise constant interpolation scheme we identify a pair, square, cube, 
etc. of $2^d$ neighbouring grid points on a grid of
level $k$ with a single grid point of the next coarsened grid 
$\Xi^{(k-1)}$. On these coarsened grids we have auxiliary fields 
$\phi_i^{(k-1)}$ representing the collective moves of the original 
fine-grained lattice. A (piecewise constant) interpolation operator 
${\cal P}$ is defined by simply adding a finite value of some variable 
$\phi_i^{(k-1)}$ on a coarse grid to each of the $2^d$ corresponding 
grid points of the next finer grid. This allows to define a Hamiltonian 
of the coarse grid in terms of the Hamiltonian on the next finer grid by
\cite{mgmc1}
\begin{equation}
 {\cal H}^{(k-1)}(\{\phi_i^{(k-1)}\}) = 
     {\cal H}^{(k)}(\{\phi_i^{(k)}\}+{\cal P}(\{\phi_i^{(k-1)}\})).
 \label{eq:gridhamiltonian} 
\end{equation}
In essence this prescription defines a Hamiltonian on the coarse grid 
$\Xi^{(k-1)}$ by freezing the field variables $\phi_i^{(k)}$ of the 
next finer grid and calculating the effect of collective
moves represented by the field variables $\phi_i^{(k-1)}$ added onto
$\Xi^{(k)}$ by the piecewise constant interpolation operator.
If the functional form of the Hamiltonian remains stable under
the coarsening prescription \cite{sokal93} this {\em multi\/}grid 
implementation of the collective move updates
minimizes the amount of computational effort compared to the 
straightforward unigrid implementation of the
collective move update, in a way similar to the Fast Fourier 
Transformation (FFT). 
Also, it allows to define the multigrid update in the following 
recursive way. Updates of level $\Xi^{(k)}$ consist of a) $n_1$
presweeps using any valid local update scheme with Hamiltonian
(\ref{eq:gridhamiltonian}), b) calculating the Hamiltonian for the next 
coarser grid $\Xi^{(k-1)}$ (which according to 
(\ref{eq:gridhamiltonian}) 
depends on the current configuration on grid $\Xi^{(k)}$) and
initializing the variables on grid $\Xi^{(k-1)}$ to zero. One then c)
updates the field variables $\phi_i^{(k-1)}$ by applying the multigrid 
update $\gamma_{k-1}$ times. To complete the update cycle one then d)
interpolates the variables of grid $\Xi^{(k-1)}$back to grid $\Xi^{(k)}$
and e) performs another $n_2$ postsweeps of the local update algorithm.
On the coarsest grid, of course, one only performs steps a) and e).
In this way we cycle through the sequence of coarsened grids 
in a specific manner which is determined by the parameters $\gamma_k$.
Particularly successful is the choice $\gamma_k \equiv 2$, a sequence 
which is commonly called W-cycle since its graphical representation
very much resembles the letter W.\footnote{See, e.g., Ref.\cite{hac85},
p.33.}

\subsection[]{Multicanonical multigrid Monte Carlo}

From the {\em uni\/}grid viewpoint it is immediately clear that
the multicanonical and multigrid methods can easily be combined for
a field-driven transition where $m=m_1$ is the average magnetization
\cite{js93b,js93c}.
Since on level $k$ we effectively always move
$2^{(n-k)d}$ spins in conjunction an accepted Metropolis move would
change the average field $m_1$ by an amount of
$2^{(n-k)d}\Delta\phi^{(k)}_{i_0}/V$.
The only modification for the update on level $k$ will therefore
be to compute the energy difference according to
\begin{equation}
     \Delta E^{{\rm muca},(k)} 
           = \Delta E^{(k)} + f(m + 2^{(n-k)d}\Delta \phi_{i_0}^{(k)}/V) - f(m),
 \label{eq:DEmucamu}
\end{equation}  
where $\Delta E^{{\rm muca},(k)}$ is the energy difference computed
with the coarse-grid Hamiltonian (\ref{eq:gridhamiltonian}) as in the 
usual canonical multigrid formulation. While this modification is obvious from
the {\em unigrid\/} point of view it should be stressed that the
modifications for a recursive {\em multigrid} implementation are
precisely the same.

In our case the doubly peaked observable $m$ which controls the
multicanonical reweighting factor is the magnetization $m_1$.
In this case the necessary modifications for a recursive multigrid update
of the multicanonical Hamiltonian ${\cal H}^{\rm muca}$ are in fact
almost trivial.
The combination of multigrid update schemes with the multicanonical
reweighting idea, however, is neither restricted
to the special choice of $m=m_1$ nor to any special choice of the
reweighting factor $f$. In general, a multigrid Monte Carlo update
of a multicanonical Hamiltonian ${\cal H}^{\rm muca}(\{\phi_i\}) = {\cal
H}(\{\phi_i\})+f(m(\{\phi_i\}))$ should be feasible and effective
whenever both the canonical Hamiltonian ${\cal H}$ {\em and} the parameter
$m$ are {\em stable under coarsening}. To see this, let us assume we
would want to reweight the canonical Hamiltonian ${\cal H}$ in
some other observable of $\{\phi_i\}$, say
$m=m_2$, rather than in $m=m_1$. In order to compute the reweighting
factor $f(m_2)$ on level $k$ we would need to know the actual value
of $m_2$ as a function of the coarse-grid variables $\phi^{(k)}_i$.
Clearly, we cannot simply compute the average value of $m_2$ by
multiplying $\Delta \phi_{i_0}^{(k)}$ with a simple factor as was
the case for the average
magnetization. Indeed, from the unigrid point of view we would need
to know the actual configuration of the original-grid variables and
the efficiency gained from the multigrid update would be lost at least
for this part of the update. It is therefore important to realize that
$m_2$ may also be calculated using the usual coarsening prescription.
In general, the analog of eq.~(\ref{eq:gridhamiltonian}) for the function
$m$ would simply read 
\begin{equation}
 m^{(k-1)}(\{\phi_i^{(k-1)}\}) = 
     m^{(k)}(\{\phi_i^{(k)}\}+{\cal P}(\{\phi_i^{(k-1)}\})).
 \label{eq:gridobservable} 
\end{equation}
Therefore we would only have to compute the coarse-grid coefficients
for the function $m^{(k-1)}$ in addition to those for ${\cal H}^{(k-1)}$
and we could then use the value of $m^{(k-1)}$ in order to 
compute the reweighting factor. From this point of view, the factors
$2^{(n-k)d}$ appearing in eq.~(\ref{eq:DEmucamu}) are nothing but the
coarse-grid coefficients for the average magnetization using piecewise
constant interpolation.

We would like to stress again at this point that the condition that
$m$ remains stable under coarsening is the {\em only} restriction
for an effective multigrid Monte Carlo update of a multicanonical
Hamiltonian.
In particular, this means that
a) the reweighting factor $f(m)$ may be computed for {\em any} function
$f$. In fact, the step functions normally employed are examples for
rather special, highly non-linear functions.
b) The condition also allows
a combination of multigrid techniques and multicanonical reweighting
if we take the canonical Hamiltonian ${\cal H}$ itself as the
observable $m$. This would be the situation for a temperature-driven
first-order transition. If ${\cal H}$ is stable under coarsening
(as we always assume it is) it therefore immediately follows that a
multicanonical multigrid Monte Carlo simulation would be perfectly
feasible in this case as well.
c) We also believe, that multicanonical multigrid Monte
Carlo simulations should be feasible for models other than those
characterized by a Hamiltonian of the form (\ref{eq:model2}) such as
non-linear O(n) sigma-models. If, e.g., one would want, for some
reason, to simulate an $XY$-model with a multicanonical reweighting
factor $f=f(s_x)$ where $s_x$ is the average $x$-component of
the spins one might express $s_x$ as $(\sum_i \cos(\Theta_i))/V$
and the latter function is stable under coarsening for piecewise
constant interpolation if we allow
for an additional term $(\sum_i \sin(\Theta_i))/V$ on the coarsened
grids. More realistic but nevertheless feasible as well would be the
case where the reweighting variable is the squared magnetization
\cite{neuhaus94}.
d) Furthermore, the fact that we only need to know the actual value
of $m$ for each coarse-grid update and the fact that this value
may be computed from eq.~(\ref{eq:gridobservable}) also entails that
the multicanonical multigrid Monte Carlo method is, in general, not 
restricted to the piecewise constant interpolation scheme.
e) Finally, it should be recalled that multigrid techniques
are sophisticated update schemes which do not presuppose
specific update algorithms. In principle, we may therefore
use any other valid update algorithm 
on the coarsened grids instead of the Metropolis update.

In this paper we will substantiate our claim that multicanonical 
multigrid Monte Carlo simulations are both feasible and profitable
by a careful analysis of the performance of the method for
the model (\ref{eq:model1}), (\ref{eq:model2}). For other situations
the method should be tested by explicit simulations. As long as
the central condition for the multicanonical multigrid approach
is fulfilled, however, we do not expect any difficulties regarding
the feasibility of the method.

\subsection[]{Effective autocorrelation time}

Before discussing our results it is worthwhile to comment
on a technical complication which arises in evaluating the
efficiency of multicanonical simulations.
In previous investigations it was the exponential autocorrelation 
time measured in the multicanonical distribution 
which was used to estimate the performance of the multicanonical
algorithm. Alternatively, autocorrelations were also measured by
counting the average number of sweeps needed to travel from one
peak maximum to the other and back (see section 4.1. below).
This nicely illustrated the absence of an
exponential slowing down 
\cite{muca1,jbk92,multimagnetical,muca2a,muca2b,muca3,muca4}.
It should be stressed, however, that neither
the exponential autocorrelation time nor the 
diffusion time\footnote{In analogy to canonical simulations this 
is often called ``tunneling time'' even though this is quite a
misleading terminology in the multicanonical case where the dynamics
is of a diffusive type.}
can a priori serve as a fair quantitative measure for comparison of
the performance of 
multicanonical simulations with canonical update schemes.
The reason is that the estimator 
$\hat{\Obs} = \sum_{i=1}^{N_m} \Obs_iw_i / \sum_{i=1}^{N_m} w_i$ 
of canonical observables (\ref{eq:mucareweight}) 
is a ratio of two different multicanonical observables which may have 
different autocorrelations {\em and}, moreover, are usually
strongly cross-correlated. It is thus not immediately
obvious how autocorrelations relevant for canonical quantities should be
defined and measured in multicanonical simulations.
For a fair comparison with canonical simulations we therefore
define \cite{js93c} an {\em effective} autocorrelation time $\tau^{\rm
eff}$ and 
write the error estimate also 
in the multicanonical case in the standard form 
\begin{equation}
     \epsilon^2_{\hat{\Obs}} = (\sigma^2_{\Obs_i})^{\rm can} 
                \frac{2\tau^{\rm eff}_{\Obs}}{N_m},
 \label{eq:taueff}
\end{equation}
where $N_m$ is the number of multicanonical measurements and 
$(\sigma^2_{\Obs_i})^{\rm can}$ is the variance of 
$\Obs_i$ in the canonical ensemble computed according to
eq.(\ref{eq:mucareweight}).
A simple and numerically stable way to obtain an estimate for the 
effective autocorrelation time $\tau^{\rm eff}_{\Obs}$ is to compute
the error $\epsilon^2_{\hat{\Obs}}$ of the estimator $\hat{\Obs}$
by jackknife blocking.
In principle, $\tau^{\rm eff}_{\Obs}$ can also  
be calculated by applying standard error propagation starting from
the basic reweighting formula (\ref{eq:mucareweight}). 
As shown in Appendix A the squared canonical error 
$\epsilon^2_{\hat{\Obs}}$
of the canonical estimator $\hat{\Obs}$ of an observable
$\langle {\cal O} \rangle_{\rm can}$ based on $N_m$ 
multicanonical measurements is then given by 
\begin{eqnarray}
    \epsilon^2_{\hat{\Obs}} = & \langle \Obs \rangle^2_{\rm can}
      & [ \frac{\langle \Obs_i w_i; \Obs_i w_i \rangle}
                {\langle \Obs_i w_i \rangle^2} 
                  \frac{2 \tau^{\rm int}_{\Obs w; \Obs w}}{N_m}
       +   \frac{\langle w_i; w_i \rangle}
                {\langle w_i \rangle^2} 
                   \frac{2 \tau^{\rm int}_{w; w}}{N_m} \nonumber \\
  & &  - 2 \frac{\langle \Obs_i w_i; w_i \rangle}
                {\langle \Obs_i w_i \rangle \langle w_i \rangle }
                \frac{2 \tau^{\rm int}_{\Obs w; w}}{N_m} ] .
 \label{eq:fullerror}
\end{eqnarray}
where $\langle x;y \rangle \equiv \langle xy \rangle - \langle x 
\rangle \langle y \rangle$ is the covariance matrix of two
observables $x$ and $y$.
From this expression it is clear that in general three different 
integrated autocorrelation times $\tau^{\rm int}_{\Obs w; \Obs w}$,
$\tau^{\rm int}_{w;w}$, 
and $\tau^{\rm int}_{\Obs w;w}$ of the observables $\Obs w$ and $w$ 
(measured in the multicanonical distribution)
must be taken into account.
%
                  \section{Results}
%
We have simulated the model (\ref{eq:model1},\ref{eq:model2}) in two 
dimensions ($d=2$) at three different points in the
($\mu^2,g$)-plane, namely we have taken $g=0.25$ and varied $\mu^2$ as
$\mu^2 = 1.30, 1.35$, and $1.40$. For this value of $g$,
the second-order phase transition to the disordered phase occurs at 
$\mu^2_c = 1.265(5)$ as it was determined in Ref. \cite{tc90}, 
confirmed in Ref. \cite{mf92}, and reproduced in our own simulation 
(see section 5). With this choice of parameters
our simulations were performed in a regime which shows the typical 
first-order behavior already for relatively small lattices.
For large linear lattice size $L$ the ratio between the
maxima and the minima of the histogram for $m_1$ will then easily take on
several orders of magnitude. 
For the severest case which we have investigated ($\mu^2=1.40, L=64$) 
this ratio, e.g., is already more than 9 orders of magnitude. 
In these extreme cases an important point of the
multicanonical algorithm is the way to obtain the trial histogram
since the performance of the multicanonical simulation strongly
depends on the quality of the assumed trial distribution.
If there is no chance to make an initial guess on the basis of
some knowledge of the system, the most straightforward way
to proceed is by iterations which in our case was done as follows.
Starting with a canonical simulation we performed some thermalizing 
sweeps and then obtained a first histogram on the basis of
$50\,000 \times n_e$ sweeps. Here $n_e$ is a parameter which allows to
adjust the time scale of the MC process, i.e., measurements were always 
taken only after every $n_e$ ``empty'' sweeps. In the multigrid case we 
count a complete cycle as one sweep, and we only performed presweeps,
i.e., we always had $n_1 = 1$ and $n_2 = 0$.
In order to maintain a roughly constant Metropolis acceptance rate of
about $50$\% we had to scale down the maximal step width 
$\propto 0.6^{n-k}$ which conforms with recent analytical investigations
of the Metropolis acceptance rate \cite{gp93}.
The first histogram was then symmetrized and any empty bins between the 
peaks were filled by interpolation using rough estimates for the 
interface tension obtained from simulations of the smaller
lattices. Using this histogram as a first guess 
to construct the reweighting factor $\exp(-f)$ we performed
another $50\,000 \times n_e$ multicanonical sweeps. The resulting
histogram proved to be sufficient for lattice sizes $L = 8, 16$, 
and $32$. To obtain high precision the resulting histogram nevertheless 
was in any case again symmetrized and taken for the final simulation run.
For $L = 64$ we have done one more iteration of $1\,000\,000 \times n_e$ 
sweeps and used only this resulting histogram as trial distribution for 
the final run. For the determination of the trial histograms, 
which always had a bin size of $0.008$, we have in 
any case used the W-cycle and, to allow for a direct comparison, 
we have taken the same trial histograms for the final runs of
both the standard multicanonical simulation and the combination 
with the multigrid scheme. In our final runs we have in each case 
performed $1\,000\,000 \times n_e$ sweeps after discarding 
$10\,000 \times n_e$ sweeps for thermalization.
To allow for later
histogramming and flexible analysis of autocorrelations we have 
recorded the time series of $K_0$, $M_1$,
$M_2$, and $M_4$, and all errors were computed by jackkniving
\cite{jackknife} the data with 50 blocks.  

Figures 2(a-c) show the flat multicanonical distributions and the 
canonical double-peak histograms of $m_1$ after reweighting for 
$\mu^2 = 1.30$ and different lattice sizes $L= 8, 16$, and $32$.
The quality of the flatness of the multicanonical distributions
for our largest lattices ($L=64$) and for $\mu^2 = 1.30, 1.35$, and
$1.40$ can be judged from Figs. 3(a-c). Although the multicanonical 
distributions are essentially flat between the peaks there still is some
structure in the multicanonical distributions which affects
autocorrelations of $m_1$. Note the flat regions in the canonical
histograms for $m_1 \approx 0$ which reflect the fact
that mixed phase configuration for some range of $m_1$ all look 
similar to the one displayed in Fig. 1. Different total
magnetizations in this regime result from a relative shift of the 
two interfaces and the free energy of these configurations is the
same as long as interactions between the interfaces are negligible.
The arrow in Fig. 3c indicates the value of $m_1$ for the configuration
displayed in Fig. 1.

The drastic difference between the canonical and the multicanonical
updates can be illustrated by looking at the time
series of $m_1$ as shown in Figs. 4 (a-d). While the time series
in the canonical case clearly displays the instantons characteristic
for tunnelling processes the observable in the multicanonical 
simulation shows a random walk like behavior. In either case the
use of the multigrid update affects the time scale of the 
autocorrelations.
Note that the time scale in the figures has been adjusted in such
a way that they display the time evolution over a length
of roughly $30 \times \tau^{\rm int}_{m_1}$ in either case 
(cp. Tables \ref{table:canonical} and \ref{table:mucatau} below).

To get a more precise view of the performance of the different
Monte Carlo schemes we have measured autocorrelation times of
the relevant observables. In order to obtain estimates for the 
integrated autocorrelation time $\tau^{\rm int}$
a common way to proceed is to cut the sum (\ref{eq:Bdef}) 
self-consistently at $k = n_{\rm cut} \times \tau^{\rm int}$
where $n_{\rm cut}$ is usually chosen to be $6$ or $8$. 
As long as $\tau^{\rm int} \approx \tau^{\rm exp}$ this method usually
gives sufficiently reliable values. If, however, $\tau^{\rm exp}$ is
appreciably larger than $\tau^{\rm int}$ this method systematically
underestimates the integrated autocorrelation time.
Let us illustrate the problem for the rather extreme case
of $\Obs = m_1 \exp(f)$, $\mu^2 = 1.40$, and $L = 64$, cp. Fig.5.
Here we find an exponential autocorrelation time $\tau^{\rm exp} = 3330(530)$
(in units of cycles) which is more than $4$ times larger than the true 
integrated autocorrelation time $\tau^{\rm int} = 778(63)$ 
(cp. Table \ref{table:taueffodd} below). 
Computing the integrated autocorrelation time by 
self-consistently cutting the sum in eq. (\ref{eq:Bdef}) therefore
underestimates $\tau^{\rm int}$ to be $505(15)$ for $n_{\rm cut} = 6$ and
$627(25)$ for $n_{\rm cut} = 8$.
In order to circumvent this problem we have therefore proceeded
as follows (see Fig. 5).
For the exponential autocorrelation time $\tau^{\rm exp}$ we have 
first obtained a rough guess $\tau^{{\rm exp},(0)}$ by a linear fit of 
$\ln A(j)$ from $j = \tau^{\rm int} \ldots 3\tau^{\rm int}$ where $\tau^{\rm
int}$ is the integrated
autocorrelation time obtained by cutting eq.(\ref{eq:Bdef}) 
self-consistently with $n_{\rm cut} = 8$. The inset in Fig. 5
shows a logarithmic plot of the autocorrelation function $A(j)$ 
for the example discussed above together with this first rough
guess shown by the dashed line. Clearly $A(j)$ does not behave
like a single exponential as would be the case if
$\tau^{\rm int} = \tau^{\rm exp}$
and consequently one must be careful to fit $\ln A(j)$ sufficiently
far away from the origin. For our first approximation we obtained
in this case $\tau^{{\rm exp},(0)} = 2480$.
We have then performed a three-parameter fit of $\tau(k)$ according to 
eq.(\ref{eq:Bfit}) in the range
$k = \tau^{{\rm exp},(0)} \ldots 3\tau^{{\rm exp},(0)}$
which yielded the values for $\tau^{\rm int}$ and $\tau^{\rm exp}$ quoted in Tables
\ref{table:canonical}-\ref{table:taueffeven}.
Fig. 5 shows $\tau(k)$ and the corresponding fit. The horizontal lines
represent our value of $\tau^{\rm int}$ together with its error bounds and
one clearly sees that $\tau(k)$ saturates towards this values
for $k \rightarrow \infty$. Again we see that the data do not yet 
saturate for $k = 6200 \approx 8\tau^{\rm int}$. The solid line in the
inset shows a linear fit of $\ln A(j)$ in the same
range $j = \tau^{{\rm exp},(0)} \ldots 3\tau^{{\rm exp},(0)}$.
This fit yields a value
$\tau^{\rm exp} = 2780(670)$ which is consistent with the one we quote but has
larger error bounds. Summing up, we note that
fitting $\tau(k)$ according to eq.(\ref{eq:Bfit}) rather than
fitting $\ln A(j)$ produces simultaneously unbiased values for
both $\tau^{\rm exp}$ and $\tau^{\rm int}$ with smaller
statistical uncertainties. 
Also these fits were satisfactorily stable against variations of the
fitting range. As usual, error bars for the values of 
$\tau^{\rm int}$ and $\tau^{\rm exp}$ obtained by this fitting procedure were obtained
using the jackknife method. 

\subsection[]{Autocorrelations in canonical simulation}

For later comparison with the multicanonical case 
we have first performed a number of canonical simulations using both 
the standard local single-hit Metropolis update and the multigrid 
W-cycle. Since the main focus of this paper, however, 
was to investigate the improvement gained
by combining the multicanonical approach with multigrid techniques,
standard canonical simulations were done only 
for $\mu^2 = 1.30$ and only for small lattices ($L = 4,8,16$). 
Table \ref{table:canonical} shows the measured autocorrelation times for 
$\Obs = m_1$ and $m_2$ (always given in units of sweeps resp. cycles).
For $\Obs = m_1$ we see that within the error bars the integrated 
autocorrelation times do not differ from the exponential 
autocorrelation times, 
i.e., in this case we are dealing with an almost purely exponential
autocorrelation function 
(see below for a theoretical explanation of this behavior).
For the even observable $\Obs = m_2$ we find
on the other hand that the exponential autocorrelation time $\tau^{\rm
exp}$ is 
appreciably larger than the integrated autocorrelation time $\tau^{\rm
int}$. The 
difference between $\tau^{\rm int}$ and $\tau^{\rm exp}$ increases on larger lattices.
Comparing the improvement of using multigrid techniques with a
W-cycle gives a factor of roughly $10 \sim 20$.
An improvement of this order had already been found in Ref.\cite{mgmc1}
for simulations at the critical line.
In either case the autocorrelation times, however, diverge 
exponentially with increasing linear lattice size $L$. 

To obtain a rough estimate of the order of magnitude of the quantities
involved we have fitted the integrated autocorrelation times of $m_1$ 
according to $\tau^{\rm int} = {\rm const} \times L^\alpha \exp(2\sigma L)$.
We find $\tau^{\rm int} = 6.41 \times L^{2.13} \exp(2\times 0.027 L)$ for the 
Metropolis case and $\tau^{\rm int} = 4.83 \times L^{0.24} \exp(2\times 0.10 L)$ 
for the W-cycle. Since we expect that the exponent should depend on
the interface tension we have also performed a constrained fit
using for $\sigma$ the value for the interface tension obtained from our
multicanonical simulations ($\sigma = 0.03443$, see below). For this 
fit we find 
$\tau^{\rm int} = 7.30(36) \times L^{2.01(22)} \exp(2\sigma L)$ for the Metropolis
case and $\tau^{\rm int} = 1.638(52) \times L^{1.366(16)} \exp(2\sigma L)$
for the W-cycle. 
Clearly, these fits can only give a rough estimate of the magnitude of 
the relevant quantities for two reasons. First we have only three data
points for fitting two resp. three parameters. 
Second, we expect corrections to be still appreciable at least for the 
smallest lattice used ($L = 4$).
Nevertheless, comparable numbers were found for the two-dimensional
Ising model where in \cite{multimagnetical} a behavior of 
$\tau^{\rm int} = 6.80 L^{2.14}\exp(2\times 0.185 \times L)$ was found for
the canonical heatbath. For the $7$-state Potts model fits of
the form $1.01 L^{2.31}\exp(2\times 0.01174 \times L)$ have been 
reported in Ref.~\cite{rummukainen93}.

The dynamical origin of the autocorrelation time for 
$m_1$ may be illustrated by a simple two-state flip model.
We measured the mean time of staying in one of the
potential wells by digitalizing the 
evolution series corresponding to a simple two-state model with single 
flip dynamics \cite{flip}. 
This procedure is illustrated for
the time series in Figs. 4a and 4c.
Counting the number of Monte Carlo time steps that the systems needs
to flip from one state to the other we measure an average 
flip rate. A theoretical analysis of this single flip dynamics 
shows that the exponential autocorrelation
time for this model is given by $\tau^{\rm flip}$ where $4\tau^{\rm flip}$ is the
average time the system needs to flip from one state to the other
and back. The measured flipping times are also shown 
in Table \ref{table:canonical}. 
For large $\tau^{\rm flip}$ the variance $\sigma^2$ of 
$\tau^{\rm flip}$ is given
by $(\tau^{\rm flip})^2$ itself as can be calculated in the single flip
dynamics and as we have verified in our canonical simulations.
The error bars for the flipping times $\tau^{\rm flip}$ therefore were 
calculated as $\delta\tau^{\rm flip} = \tau^{\rm flip}/\sqrt{n_e}$ where
$1/n_{\rm flip}$ is the measured total flip rate, i.e.
$2\tau^{\rm flip}n_{\rm flip}=N_m \times n_e$. As can be seen in 
Table \ref{table:canonical} application of the multigrid 
algorithm speeds up the Monte Carlo process by increasing the average
flip rate by a roughly constant factor.

\subsection[]{Autocorrelations in multicanonical simulation}

In the multicanonical simulation the exponential supercritical
slowing down is overcome by simulating an auxiliary distribution 
which is flat between the two maxima of the histogram, 
cp. Figs. 2 and 3, and in this case we expect 
the autocorrelations to be governed by some random walk dynamics,
cp. Figs. 4b and 4d.
Since multigrid techniques can be applied in the multicanonical
distribution as well it is of interest to see whether a further
reduction of autocorrelations can be achieved by this combination.
From the rather different scales of Figs. 4b and 4d it is already clear
qualitatively that autocorrelations are in fact reduced.
In order to see quantitatively how {\em multicanonical} simulations
are improved by multigrid updating we first measured autocorrelations 
of the corresponding observables in the multicanonical distribution 
using both the standard single-hit Metropolis algorithm and the W-cycle.
Table \ref{table:mucatau} shows our results for this case.
We see that for $\Obs = m_1$ it is again found that 
$\tau^{\rm int}_{m_1} \approx \tau^{\rm exp}_{m_1}$, i.e., also in the
multicanonical dynamics the 
autocorrelation function decays like a pure exponential. For the even 
observable $\Obs = m_2$, on the other hand, there is again a difference 
between integrated and exponential autocorrelation times. 

Comparing the absolute values for the Metropolis update and for the
W-cycle we find that for both observables the multigrid method reduces
the autocorrelation times by a factor of roughly $15 \approx 20$.

We have also 
looked at the lattice size dependence of the autocorrelation times by
fitting the data for $\Obs = m_1$ according to
$\tau^{\rm int} = a_{\rm int} L^{z_{\rm int}}$ or 
$\tau^{\rm exp} = a_{\rm exp} L^{z_{\rm exp}}$ (where $z=\alpha d$).
Here we first note that trivially the exponent $z_{\rm int}$ for the
integrated autocorrelation times agrees with the exponent $z_{\rm exp}$
for the exponential autocorrelation times. We also find that fitting
only the data for $L=8,16$, and $32$ yields approximately the same 
exponents for the Metropolis case and for the multigrid W-cycle.
Looking at the dependence on the parameter $\mu^2$ we 
find that the exponent increases with $\mu^2$, i.e. with the strength 
of the transition. 
Fitting the integrated autocorrelation times obtained from 
the multigrid simulation for lattice sizes, $L=16,32$, and $64$, we 
obtain exponents of about $2.2$, $2.5$, and $2.7$ for 
$\mu^2 = 1.30$, $1.35$, and $1.40$. Including the $L=8$-data worsens 
the fits, and we therefore believe that for a reliable 
estimate one would need to include even larger lattices.
In general, we observe that the fits have rather large chi-squares, and 
we hesitate to draw any definite conclusions. The deviations from 
linearity found in the log-log-fits are believed to be effected by the 
fact that the multicanonical histograms are not ideally flat. In Figs.
2 (a-c) and 3 (a-c) the multicanonical histograms look better than
they actually are as a result of the logarithmic scale. On a linear scale
one still discerns some structure in the supposedly flat region 
between the peaks at least for the larger lattices. Therefore
the statistical accuracy of our data seems to be better than the 
systematic fluctuations of the autocorrelation times caused by
imperfect multicanonical trial histograms.

With respect to the purely multicanonical dynamics, 
we conclude that the multigrid technique does not affect the exponent 
$z=\alpha d$. However, it is by largely reducing the overall scale of the 
autocorrelation, i.e. by reducing the prefactor $a$, that
application of multigrid techniques gives an improvement factor
of roughly $15 \sim 20$, i.e. of about one order of magnitude. 
The improvement factor shows a weak tendency to increase if $\mu^2$
approaches the critical value.

\subsection[]{Effective canonical autocorrelations in multicanonical 
              simulation}

Clearly, the multicanonical reweighting factor is an algorithmical 
artefact introduced in order to obtain higher statistical accuracy for 
the measurement of canonical observables. 
For a fair comparison between the canonical and the multicanonical 
simulation we therefore have to estimate the error bars associated
with the {\em canonical} observables.

\paragraph[]{ Odd observables: }
For odd observables standard error 
propagation starting from eq.(\ref{eq:mucareweight}) shows that 
the effective autocorrelation time $\tau^{\rm eff}_{\Obs}$ 
for canonical estimates obtained by 
measurements in the multicanonical distribution is given by 
\begin{equation}
     \tau_{\Obs}^{\rm eff} = 
          \frac{\sigma^2_{\hat{\Obs}}}{(\sigma_{\Obs_i}^2)^{\rm can}}
            \tau_{\Obs w; \Obs w},
 \label{eq:taueffodd}
\end{equation}
(see Appendix A) with the effective multicanonical variance
\begin{equation}
     \sigma^2_{\hat{\Obs}} = \langle \Obs \rangle^2_{\rm can} 
         \frac{\langle \Obs_iw_i ; \Obs_iw_i \rangle}
              {\langle \Obs_iw_i \rangle^2}
              = \frac{\langle \Obs_iw_i;\Obs_iw_i \rangle}
                     {\langle w_i \rangle^2}.
\label{eq:sigmamucaodd}
\end{equation}
Table \ref{table:taueffodd} shows the measured integrated and 
exponential autocorrelation times $\tau^{\rm int}$ and $\tau^{\rm exp}$ for 
$\Obs = m_1 \exp(f)$. Also shown are
the effective multicanonical variance $\sigma^2_{\hat\Obs}$ according to 
eq.(\ref{eq:sigmamucaodd}),
the canonical variance $(\sigma^2_{\Obs_i})^{\rm can}$ of $\Obs = m_1$, 
which can be computed in a multicanonical simulation by using 
eq.(\ref{eq:mucareweight}),
the effective autocorrelation time $\tau^{\rm eff}_{m_1}$
computed according to eq.(\ref{eq:taueffodd}), and the
diffusion time $\tau^{\rm flip}_{m_1}$ defined in analogy to section
4.1.

First, we notice that, within error bounds, the exponential 
autocorrelation times for $\Obs = m_1 \exp(f)$ agree 
with the purely multicanonical autocorrelation times for $\Obs = m_1$ 
listed in Table \ref{table:mucatau}. This is not surprising 
since we are still dealing with an odd observable whose slowest
autocorrelation mode should be same.
The integrated autocorrelation times on the other hand in this case
differ appreciably. This is an indication that the autocorrelation
function (\ref{eq:A(j)})
does not behave like a simple exponential $\propto \exp(-j/\tau^{\rm
exp})$.
Rather it is composed of many different modes with only the slowest mode
decaying with $\tau^{\rm exp}$ as illustrated in the inset of Fig. 5.
The relative difference between the integrated and the exponential
autocorrelation time increases both with the size of the
system and with $\mu^2$. The ratio does not depend on the other
hand on the use of the algorithm, being roughly the same both for 
the standard Metropolis update and for the multigrid update. 

Table \ref{table:taueffodd} lists both the effective
multicanonical and the canonical variances.
These allow to compute the final 
effective autocorrelation times which are also reported. 
While the canonical variance depends only weakly on the size
of the system the effective multicanonical variance 
$\sigma^2_{\hat{\Obs}}$ varies appreciably with the
linear lattice size $L$. In the worst case, $\mu^2 = 1.40$ and
$L=64$ the ratio is already 
$\sigma^2_{\hat{\Obs}}/(\sigma^2_{\Obs_i})^{\rm can} \approx 11$.
Consequently the effective autocorrelation time which should be used for
comparisons with canonical algorithms is much larger than the simple 
exponential autocorrelation time $\tau^{\rm exp}$.

To allow for further comparison with the literature, 
we have looked also for this situation at the exponents $z_{\rm int}$ 
resp. $z_{\rm exp}$ of the
power-like divergence $\tau = a L^z$. In contrast to the
purely multicanonical case the exponents $z_{\rm int}$ here differ
from the exponents $z_{\rm exp}$. While the exponents $z_{\rm exp}$ for 
$\Obs = m_1 \exp(f)$ roughly agree with those for the purely 
multicanonical observable $\Obs = m_1$ and thus increase with $\mu^2$, 
the exponents 
$z_{\rm int}$ seem to stay constant with increasing $\mu^2$. Finally we note
that the exponent $z_{\rm eff}$ for the effective autocorrelation
time $\tau^{\rm eff}$ again increases with $\mu^2$ which directly 
reflects the scaling of the ratio
$\sigma^2_{\hat{\Obs}}/(\sigma^2_{\Obs_i})^{\rm can}$
with $\mu^2$ and $L$. The increase of $z_{\rm eff}$ with $\mu^2$
as compared to $z_{\rm int}$ is illustrated in Fig. 6. 
Here the integrated autocorrelation times $\tau^{\rm int}$ for
$\Obs = m_1\exp(f)$ 
are shown together with the corresponding effective autocorrelation 
times $\tau_{\rm eff}$.
The straight lines show fits of the form
$\tau^{\rm int} = a_{\rm int} L^{z_{\rm int}}$ and 
$\tau^{\rm eff} = a_{\rm eff} L^{z_{\rm eff}}$. From this figure it
can also be seen that the data for the smallest lattice, $L=8$, need
still be excluded to obtain satisfactory fits. Fitting the data for 
$L=16,32$, and $64$ we find for the effective exponents $z_{\rm eff}$
values of about $2.3$, $2.7$, and $3.0$ for $\mu^2 = 1.30$, $1.35$, and
$1.40$.
The exponents obtained in our work confirm qualitatively the exponents
found for standard multicanonical simulations of the two-dimensional
$q$-state Potts model where exponents of $z = 2.65(5)$ for $q=7$ 
\cite{jbk92} and $z = 2.65(2)$ for $q=10$ \cite{muca1} have been
obtained from analyses of the diffusion times.
Note that for a random walk like behavior as in the multicanonical
case one cannot unambiguously identify distinct states anymore. 
One often employed possibility is to measure the average number of
multicanonical sweeps or multigrid cycles needed to travel
from one (canonical) peak maximum to the other and back.
In analogy to the definition for canonical simulations
the $\tau^{\rm flip}_{m_1}$ given in Table \ref{table:taueffodd}
are one quarter of this average travel time. By using this
definition of $\tau^{\rm flip}_{m_1}$ we obtain a nice agreement
with $\tau^{\rm eff}_{m_1}$ at least for the large lattices.
A priori, however, other definitions of $\tau^{\rm flip}$ are
reasonable as well (e.g., using $\langle |m_1| \rangle_{\rm can}$
for the cuts instead of the peak locations),
and it is difficult to argue which one should give the best quantitative
agreement  with the unambiguously defined effective autocorrelation time
$\tau^{\rm eff}$. For this reason a direct measurement of $\tau^{\rm
eff}$ is to be preferred rather than any analogue of the two-state
flip model.

For odd observables the distinction between the directly measured 
integrated autocorrelation time and the effective autocorrelation 
time does not pertain to the comparison between the standard 
multicanonical Metropolis update and combination of the 
multicanonical approach with multigrid techniques. 
Since the multicanonical approach is a mere reweighting technique
$\sigma^2_{\hat{\Obs}}$ and $(\sigma^2_{\Obs_i})^{\rm can}$
are not affected by applying different update algorithms. Hence
we find indeed that the same improvement 
factors of about $15 \sim 20$ are gained both for the integrated 
autocorrelations and for the effective autocorrelation 
times.\footnote{Apart form work estimates to be discussed below.}    
These effective improvement factors are slightly smaller than those 
found for $\Obs = m_1$ from Table \ref{table:mucatau} and again show
a tendency to increase when $\mu^2$ approaches the critical value.
\paragraph[]{ Even observables: }
For even observables we cannot exploit the fact that 
$\langle \Obs_i w_i \rangle$ vanishes identically for reasons
of symmetry in order to simplify the error propagation formula
(\ref{eq:fullerror}). In general, to obtain estimates for the canonical 
error of an observable $\Obs$ we therefore have to take recourse to the
full expression of error propagation (\ref{eq:fullerror}).
Table \ref{table:taueffeven} shows our results for the even 
observable $\Obs = m_2$ from our simulations at $\mu^2 = 1.30$.
Here we list measured values for all quantities which enter the
error propagation formula (\ref{eq:fullerror}). 
The mean values, variances, and covariance, for the Metropolis
case and for the W-cycle, are consistent within error bars, as, of course,
should be the case since these quantities do not depend on the update
algorithm. 
The integrated autocorrelation times, however, again differ by
a factor of roughly $20$. We did not list the corresponding
exponential autocorrelation times since these agree, within error
bounds, with the exponential autocorrelation times for $\Obs = m_2$
listed in Table \ref{table:mucatau}.
We have also checked that
$\tau^{\rm int}_{\Obs w; w} \approx \tau^{\rm int}_{w; \Obs w}$
as would be expected because of time reversal invariance.

Next to these values we then list in Table \ref{table:taueffeven} 
the squared statistical error $\epsilon^2$ for the canonical estimator 
of $\Obs = m_2$ calculated by error propagation from
the data listed before. Note that the error of this error, however, as 
well as {\em all} the errors given in the Table 
were not computed by error propagation but, as usual, calculated 
directly by jackkniving.

Clearly, it is this statistical error for the canonical estimates
of the observable which one wants to reduce by sophisticated
Monte Carlo methods. When interpreting the statistical errors 
reported in Table \ref{table:taueffeven} the time scale set by $n_e$ 
should also be taken into consideration. While the squared errors 
$\epsilon^2$
are approximately of the same order for the Metropolis (M) update and 
for the multigrid W-cycle update (W) we also had to perform many more
Metropolis updates since we had adjusted $n_e$. If, e.g., for
$L=8$ the statistical error for the Metropolis update is only twice as 
large as the one for the multigrid update, we also had $n_e = 5$(M) 
resp. 1(W), cp. Table \ref{table:mucatau}. Therefore the improvement is
given by $(0.540/0.2442)\times 5 \approx 11$ which is roughly of the 
same size as the ratio of the measured autocorrelation times.

Another technical remark is due at this point.
Applying formula (\ref{eq:fullerror}) to calculate the canonical
error of multicanonical measurements we run into a nasty problem
of numerical cancellation.
To illustrate this cancellation problem let us look at the data
for $L=64$. Here we find for the first two
terms in eq.(\ref{eq:fullerror})
\begin{equation}
     \frac{ \langle \Obs_i w_i; \Obs_i w_i \rangle  }
          { \langle \Obs_i w_i \rangle^2 } \tau^{\rm int}_{\Obs w; \Obs w}
    +\frac{ \langle w_i; w_i \rangle  }
          { \langle w_i \rangle^2 } \tau^{\rm int}_{w; w}
    = 1006.22
\end{equation}
and for the third term
\begin{equation}
    2 \frac{ \langle \Obs_i w_i; w_i \rangle  }
          { \langle \Obs_iw_i \rangle \langle w_i \rangle } 
      \tau^{\rm int}_{\Obs w; \Obs w}
    = 1006.32,
\end{equation}
i.e., we have a numerical cancellation up to the fifth digit.
Also if we look at $\sigma^2_{\hat{\Obs}}$ we find the same problem
which should not come as a surprise if we recall that in the definition
(\ref{eq:defmucasigma}) of $\sigma^2_{\hat{\Obs}}$
we simply dropped the $\tau$'s in eq.(\ref{eq:fullerror}). 
Since from Table \ref{table:taueffeven} we see that
for the even observable $m_2$ we always have
$\tau^{\rm int}_{\Obs w;\Obs w} \approx \tau^{\rm int}_{w;w} \approx \tau^{\rm int}_{\Obs w;w}$
the numerical cancellation should therefore carry over to 
$\sigma^2_{\hat{\Obs}}$ as well.
Consequently, we obtain numerical results for the statistical error 
estimate which may be completely erroneous. In fact, for $L = 64$
the effective autocorrelation time turns out to
be negative which, of course, is complete boloney.
Therefore it is somewhat difficult to judge the quality of the 
performance of the multicanonical simulation of even canonical 
observables by applying error propagation.
Alternatively we can, of course, judge the improvement gained by applying
multigrid techniques by comparing the errors obtained by 
jack-knife blocking procedures.
For comparison we therefore have listed the squared canonical
errors obtained in this manner as well as the effective autocorrelation 
times derived from these jackknife errors. 
These values in general turn out to agree roughly with the 
calculated errors for small lattices but deviate strongly for 
our large lattices. In general we tend to believe that in this
case the error estimates obtained by direct jackkniving are
more reliable than the ones calculated by error propagation.

Finally it should be remarked that a measurement of even 
observables in the multicanonical distribution is somewhat academic
anyway since they may already be measured quite accurately in the 
canonical distribution. Comparing the autocorrelation times given in 
Table \ref{table:taueffeven} with the autocorrelation times for the
canonical simulation reported in Table \ref{table:canonical} we find
that the autocorrelation times are roughly of the same order
of magnitude, and may even become larger by multicanonical 
sampling. In fact, for $\Obs=m_2$ multicanonical sampling only increases
the statistics in the exponentially suppressed tail of the canonical 
probability distribution $P(m_2)$.
This observation, however, does not affect our overall claim that 
the combination of multigrid techniques with multicanonical updating
does significantly enhance the performance of the Monte Carlo process
even though this improvement is practically irrelevant in the case
of even observables.

\subsection[]{Real-time performance}

To conclude the analysis of the performance of the multicanonical
multigrid algorithm we finally need to look at the real-time work 
needed for the different algorithms.
From a theoretical work estimate \cite{mgmc1} it follows that for 
$\gamma < 2^d$ the additional work necessary to perform one complete
W-cycle in comparison to a simple multicanonical sweep is given by a
constant factor. For $\gamma = d = 2$ this factor is predicted to
be close to 2. With our implementation on a CRAY X-MP we have measured
updates times per site and cycle of 
$t_{\rm real} = 11.2, 10.3, 9.5, 9.1 \mu s$ for the W-cycle (W) resp.
$t_{\rm real} = 4.0, 3.9, 3.9, 3.8 \mu s$ for the Metropolis (M) 
algorithm for lattices of size $L=8,16,32,64$. On a $128^2$-lattice our 
program would run with $8.2$ $\mu s$ (W) resp. $3.7$ $\mu s$ (M), and 
on a $256^2$-lattice with $8.1$ $\mu s$ (W) resp. $3.7$ $\mu s$ (M).
It goes without saying that these
numbers strongly depend on hardware features of the computer and on
details of the implementation. We conclude that the gain in reduction
of the autocorrelation times of a factor of 20 is roughly halved by the 
additional work needed to perform a W-cycle. Thus it is established
that the combination with multigrid techniques enhances the
performance of the standard multicanonical algorithm by about one
order of magnitude, asymptotically independent of the linear lattice 
size $L$.

    \section{Interface tension}

Having tested the performance of the algorithms we now turn to
the evaluation of some observables of interest. Before doing so we 
recall that standard reweighting techniques \cite{histogramming}
allow to compute expectation values of observables for
an appreciable range away from the simulation point.
Since in the multicanonical case several different reweighting
factors are employed we briefly review the histogramming technique
for this case.

In order to reweight to a new set of parameters we use the notation 
of eqs.(\ref{eq:notation2},\ref{eq:notation1}) and notice that 
expectation values of 
canonical observables $\Obs = \Obs(K_0,M_1,M_2,M_4)$
are obtained from the multicanonical distribution by computing
\begin{eqnarray}
  \langle \Obs \rangle_{\rm can} (\mu^2,g) &=&  
   \frac{ \int dK_0 dM_1 dM_2 dM_4 \; \Obs e^{f(M_1/V)}
   N \exp\{-{\cal H}^{\rm muca}_{\mu^2,g} \}}
        { \int dK_0 dM_1 dM_2 dM_4 \; e^{f(M_1/V)}
   N \exp\{-{\cal H}^{\rm muca}_{\mu^2,g}  \}}
        \\
  &=& \frac{\int dK_0 dM_1 dM_2 dM_4 \; P^{\rm muca}_{\mu^2,g}
            \Obs e^{f(M_1/V)}}
           {\int dK_0 dM_1 dM_2 dM_4 \; P^{\rm muca}_{\mu^2,g}
            e^{f(M_1/V)}}
\end{eqnarray}
Here $N=N(K_0, M_1, M_2, M_3)$ denotes the density of states for
the variables $K_0$ and $M_i$,
${\cal H}^{\rm muca}_{\mu^2,g}  (K_0, M_1, M_2, M_4) 
= K_0/2 - (\mu^2/2) M_2 + g M_4 + f(M_1/V)$ 
is the multicanonical energy, and 
$P^{\rm muca}_{\mu^2,g} \propto N\exp\{-{\cal H}^{\rm muca}_{\mu^2,g}\}$ 
is the multicanonical probability distribution. Also we have dropped 
cancelling normalization factors. In a multicanonical Monte Carlo
simulation configurations are sampled with a probability
$\propto P^{\rm muca}_{\mu^2,g}$. Hence if we record the evolution 
series $M_i$ of a simulation performed for one set of parameters 
$(\mu^2,g)$ the expectation value of an observable $\Obs$ for some 
other set of parameters $(\mu'^2,g')$ can now in principle be 
calculated by multiplying with a reweighting factor. For example, the 
expectation value for $\mu'^2 \neq \mu^2$ would simply be given by
\begin{equation}
  \langle \Obs \rangle_{\rm can}(\mu'^2,g) =
    \frac{\int dK_0 dM_1 dM_2 dM_4 \; P^{\rm muca}_{\mu^2,g}
       \Obs e^{f(M_1/V)} e^{\frac{\mu'^2-\mu^2}{2}M_2}}
           {\int dK_0 dM_1 dM_2 dM_4 \; P^{\rm muca}_{\mu^2,g}
            e^{f(M_1/V)}e^{\frac{\mu'^2-\mu^2}{2}M_2}}.
\end{equation}
The only restriction for the reweighting procedure is given
by the fact that the statistical accuracy of the data deteriorates
if one reweights the data to a set of parameters far away from
the simulation point. The problem is illustrated in Figs. 7(a-f).
Fig. 7a shows the joint probability distribution $P(m_1,m_2)$
for the multicanonical simulation at $\mu^2 = 1.40$ and $L = 64$, 
and Fig. 7b shows the same distribution after reweighting to the 
canonical case. These figures are directly comparable to Fig. 3c. 
Again we see in Fig. 7a the flat region between
the peaks which allows the system to travel from states of negative to 
positive magnetization. Note that the histogram depicted 
in Fig. 7a does produce the flat one-parameter histogram of Fig. 3c
after integration over $m_2$. For the histogram reweighting, however, 
it is important to realize that also for the multicanonical situation of
Fig. 7a a reweighting in the parameter $\mu^2$ shifts the histogram
towards regions of smaller $m_2$ where the multicanonical statistics is 
as bad as a canonical simulation would be. After reweighting to 
$\mu^2 = 1.375$ and $1.35$ the resulting distributions are depicted in 
Fig. 7c resp. 7e for the multicanonical and in Fig. 7d resp. 
7f for the canonical case. Comparing the multicanonical distributions
of Figs. 7c and 7e with the original smooth distribution of Fig. 7a
one clearly sees that the histograms get increasingly noisy since the 
reweighting procedure suppresses the high statistics regions in Fig. 7a 
in favor of regions where only few configurations were sampled.
Note that the normalization was adjusted in such a way as to show the 
peaks at same height. Consequently the $z$-scale varies over
many orders of magnitude in Figs. 7 (a-f), i.e., the maxima
of the distributions vary as $4358$ (7a), $3858$ (7b),
$11941$ (7c), $282$ (7d), $678\,568$ (7e), and $130$ (7f). 
Although from Fig. 7e one would expect that the reweighting
already breaks down it was nevertheless possible to find overlapping 
regions when reweighting our data in the intervals between the
simulation points.

Since we were mainly interested in the first-order phase transition 
we did not focus on thermodynamic properties at criticality. 
We only mention that the reweighting technique in principle allows 
us to compute the susceptibility $\chi$ and specific heat $C$ at the 
second-order transition line starting from our simulation data for 
$\mu^2 = 1.30$. In this way we have determined the transition point by 
extrapolating the finite lattice peak locations of $\chi$ and $C$ for 
$L \rightarrow \infty$ and found a critical value of 
$\mu^2_c = 1.270(7)$ which is slightly larger than but still compatible 
with the value of $\mu^2_c = 1.265(5)$ found by Toral and Chakrabarti 
\cite{tc90}.

More interesting in our context of an investigation of first-order 
phase transitions is a study of the interface tension. 
Here again we may use reweighting techniques. 
As discussed in section 2 the interface tension $\sigma_L$
can easily be extracted from a histogram of $m_1$ by the relation
\begin{equation}
    \sigma_L = \frac{1}{2L} \ln \frac{P^{\rm max}}{P^{\rm min}}.
\end{equation} 
Since in the canonical distribution $P^{\rm max}$ is larger than 
$P^{\rm min}$ by many orders of magnitude a reliable numerical 
evaluation of this relation is only possible for multicanonical 
simulations. In a canonical simulation there would only be very few 
configurations (if any) around $P^{\rm min}$ and the relative 
statistical error of $P^{\rm min}$ would be prohibitively large.
Due to the flat multicanonical histograms on the other hand the region
around $P^{\rm min}$ is sampled with the same statistical
accuracy as the region around the maxima. A simple determination 
of the maximum resp. minimum of the histogram strongly depends 
on the bin size of the histogram and tends to overestimate the ratio 
$P^{\rm max}/P^{\rm min}$ for moderate bin sizes.
To avoid this problem we determined $P^{\rm max}$ and $P^{\rm min}$
by fitting parabolas to the extremal points of the histogram.
For the histograms we used a bin size of 0.004, i.e., we had
of the order of $10^3$ entries in the bins between the maxima. 
For the fits of the maxima we cut the data at 
$0.85 \times P^{\rm max}$, and for
the fits of the minima we used data from $m_1 = -0.2 \dots 0.2$.
We have checked that the results did not sensibly depend on the
specific choice of the histogramming bin size or the cutting 
parameters for the fits. 

Fig. 8 shows the interface tension $\sigma_L$ for various lattice
sizes $L$. The solid circles show the points where the actual
simulations were performed, the interpolating lines were obtained
by reweighting. Note that we have reweighted the data up to
the mid points where the reweighted data from above meet those
which were reweighted from below. Judging from Figs. 7(a-d) we believe
that this range still gives reliable values. For our small
lattices $L = 8,16$, and $32$ we were also able to reweight our data 
well beyond the critical value $\mu^2_c$ for the infinite system.
To obtain values for the infinite volume interface tension 
$\sigma_{\infty}$ we have extrapolated the (reweighted) data 
according to a fit of the form \cite{wiese92}
$\sigma_L = \sigma_{\infty} + a/L$. 
The squares in Fig. 8 show our infinite volume interface tensions
at the simulation points. The precise values are listed in 
Table \ref{table:sigma}.
 
From the universality with the two-dimensional Ising model we 
expect that the interface tension varies linearly with $\mu^2$ since 
for the Ising model the critical exponent $\nu$ is equal to $1$.
Looking at the dependence of $\sigma_{\infty}$ with $\mu^2$ 
we find indeed that the interface tension $\sigma_{\infty}$ behaves like 
$\sigma_{\infty} = a \times (\mu^2-\mu_c^2)$. 
A linear fit of the three data points of $\sigma_{\infty}$
intersects the $\mu^2$-axis at a value $\mu^2_c = 1.274(3)$ 
which agrees
with our value obtained from extrapolating the maxima of the
susceptibility $\chi$ and the specific heat $C$.
For the interpretation of these data we would like to point out, however,
that the goodness of the fit 
$\sigma_L = \sigma_{\infty} + a/L$, which is perfectly satisfactory
for large $\mu^2$, somewhat deteriorates as one approaches
the critical line. In fact, for $\mu^2 = 1.30$ a fit of the form
$\sigma_L = \sigma_{\infty} + a/L + b/L^2$ gives a better chi-squared.
Applying this fit to values of $\sigma_L$ reweighted to
values of $\mu^2$ larger than $1.31$ on the other hand does not give a 
more consistent fit. The reason for this is probably the
fact that for $\mu^2=1.30$ our histograms do not show a really flat
region around $m_1\approx 0$ yet (cp. Fig.~2a).
Hence interactions between the interfaces
apparently are not yet completely negligible.
It should also be kept in mind that in the determination of 
$\sigma_{\infty}$ in the vicinity of $\mu^2_c$ quite a bit 
of numerical analysis is involved. Our extrapolation of the infinite 
volume interface tension $\sigma_{\infty}$ to $\mu^2_c$ is therefore
to be taken with a cautious mind. In particular, our data do not
allow to decide how far away from the critical value $\mu^2_c$ the
assumed linearity of $\sigma_{\infty} = a \times (\mu^2-\mu_c^2)$
actually holds. 
%
       \section{Concluding remarks}
%
We have shown that a combination of the multicanonical reweighting 
algorithm with multigrid update techniques reduces autocorrelation times 
of the Monte Carlo process at the field-driven first-order phase
transitions of the two-dimensional $\phi^4$-model
by a factor of $\approx 20$ when compared with standard multicanonical
Metropolis updating. Taking into
account the additional work required for the multigrid W-cycle this
effectively improves the real-time performance
of the Monte Carlo process by about one order of magnitude 
compared with standard multicanonical simulations.

Having established this gain in performance 
it would now be interesting to perform simulations of the $\phi^4$-model
in three or four dimensions as the immediate next step.
Due to the generality of both the multicanonical formulation as well
as the multigrid technique the algorithm
is not restricted to only this one model and it is hoped
that the method may further enhance Monte Carlo studies
of first-order phase transitions or tunneling phenomena
in quantum statistics \cite{js93b,js93c}.
%
       \section*{Acknowledgement}
%
W.J. thanks the Deutsche Forschungsgemeinschaft for a Heisenberg
fellowship.
%
                        \appendix
              \setcounter{equation}{0} 
              \renewcommand{\theequation}{A\arabic{equation}}
              \renewcommand{\thesection}{Appendix \Alph{section}:}
%
   \section{Error propagation for reweighting simulation data}

For any observable $\Obs$ 
(e.g. $m_1 = \frac{1}{V}\sum_{i=1}^{V} \phi_i$)
expectation values in the {\em canonical } ensemble, 
$\langle \Obs \rangle_{\rm can}$, are calculated as
\begin{equation}
      \langle \Obs \rangle_{\rm can} = 
      \frac{\langle\Obs w\rangle}{\langle w \rangle} ,
\end{equation}
where $\langle \dots \rangle$ (without subscript) denote 
expectation values with respect to the {\em multicanonical }
distribution and $w = \exp(f)$ is the inverse reweighting
factor.%
\footnote{Of course, the same considerations apply to the standard
reweighting method\cite{histogramming} as well.}
In a Monte Carlo simulation with a total 
number of $N_m$ measurements these values are,
as usual, estimated by the mean values
\begin{eqnarray}
      \langle\Obs w\rangle \approx \overline{\Obs w} & \equiv &
        \frac{1}{N_m}\sum_{i=1}^{N_m} \Obs_i w_i , \\    
      \langle w \rangle        \approx \overline{w}         & \equiv &
        \frac{1}{N_m}\sum_{i=1}^{N_m} w_i ,
\end{eqnarray}
where $\Obs_i$ and $w_i$ denote the measurements for the $i$-th
configuration. 
Hence $\langle \Obs \rangle_{\rm can}$ is estimated by 
\begin{equation}
      \langle \Obs \rangle_{\rm can} \approx \hat{\Obs}
        \equiv \frac{\overline{\Obs w}}{\overline{w}} .
\end{equation}
The estimator $\hat{\Obs}$ is biased,
\begin{equation}
    \langle \hat{\Obs} \rangle = \langle \Obs \rangle_{\rm can}
     [ 1 - 
           \frac{\langle \overline{\Obs w};\overline{w} \rangle}
      {\langle \overline{\Obs w} \rangle \langle \overline{w} \rangle}
        +  \frac{\langle \overline{w}     ;\overline{w} \rangle}
      {\langle \overline{w}      \rangle \langle \overline{w} \rangle}
        + \cdots ],
\end{equation}
and fluctuates around $\langle \hat{\Obs} \rangle$ with variance,
i.e. squared statistical error
\begin{equation}
    \epsilon^2_{\hat{\Obs}} = \langle \Obs \rangle^2_{\rm can}
    [     \frac{\langle \overline{\Obs w};\overline{\Obs w} \rangle}
                {\langle \overline{\Obs w} \rangle^2}
       +   \frac{\langle \overline{w};\overline{w} \rangle}
                {\langle \overline{w} \rangle^2}
       - 2 \frac{\langle \overline{\Obs w};\overline{w} \rangle}
                {\langle \overline{\Obs w} \rangle \langle \overline{w} \rangle }
       + \cdots ] .
\end{equation}
Here $\langle \overline{\Obs w};\overline{w} \rangle \equiv 
\langle \overline{\Obs w}\overline{w} \rangle - \langle \overline{\Obs w} \rangle
\langle \overline{w} \rangle$, etc. denote (connected) correlations of the 
mean values, which can be computed as
\begin{equation}
    \langle \overline{\Obs w};\overline{w} \rangle = 
       \langle \Obs_i w_i;w_i \rangle \frac{2 \tau^{\rm int}_{\Obs w; w}}{N_m},
\end{equation}
where
\begin{equation}
    \tau^{\rm int}_{\Obs w; w} = \tau^{\rm int}_{w; \Obs w} = 
       \frac{1}{2} + 
       \sum_{k=1}^{N_m} \frac{\langle \Obs_0 w_0;w_k \rangle} 
                             {\langle \Obs_0 w_0;w_0 \rangle}
        ( 1 - \frac{k}{N_m} ) 
\end{equation}
is the associated  integrated autocorrelation time of measurements
in the multicanonical distribution.

Hence the statistical error is given by
\begin{eqnarray}
    \epsilon^2_{\hat{\Obs}} = & \langle \Obs \rangle^2_{\rm can}
    [  &  \frac{\langle \Obs_i w_i; \Obs_i w_i \rangle}
                {\langle \Obs_i w_i \rangle^2} 
                  \frac{2 \tau^{\rm int}_{\Obs w; \Obs w}}{N_m}
       +   \frac{\langle w_i; w_i \rangle}
                {\langle w_i \rangle^2} 
                   \frac{2 \tau^{\rm int}_{w; w}}{N_m} \nonumber \\
  & &  - 2 \frac{\langle \Obs_i w_i; w_i \rangle}
                {\langle \Obs_i w_i \rangle \langle w_i \rangle }
                \frac{2 \tau^{\rm int}_{\Obs w; w}}{N_m} ] .
 \label{eq:fullepsilon}
\end{eqnarray}
Since for uncorrelated measurements 
$\tau^{\rm int}_{\Obs w; \Obs w} = \tau^{\rm int}_{\Obs w; w} = \tau^{\rm
int}_{w; w} = 1/2$
it is useful to define an {\em effective} multicanonical 
variance\footnote{
      In the multicanonical distribution this is nothing but an 
      {\em abbreviation} of the expression on the r.h.s. but {\em not}
      the variance in the multicanonical distribution.}
\begin{equation}
    \sigma^2_{\hat{\Obs}} = \langle \Obs \rangle^2_{\rm can}
    [     \frac{\langle \Obs_i w_i; \Obs_i w_i \rangle}
                {\langle \Obs_i w_i \rangle^2}
       +   \frac{\langle w_i; w_i \rangle}
                {\langle w_i \rangle^2}
       - 2 \frac{\langle \Obs_i w_i; w_i \rangle}
                {\langle \Obs_i w_i \rangle \langle w_i \rangle } ],
  \label{eq:defmucasigma}
\end{equation}
such that the error (\ref{eq:fullepsilon}) can be written in the usual 
form
\begin{equation}
    \epsilon^2_{\hat{\Obs}} \equiv \sigma^2_{\hat{\Obs}} 
                   \frac{ 2 \tau_{\Obs}}{N_m},
\end{equation}
with $\tau_{\Obs}$ collecting the various autocorrelation times in an 
averaged sense.
For a comparison with canonical simulations we need one further step 
since
\begin{eqnarray}
     ( \epsilon^2_{\hat{\Obs}} )^{\rm can}  
      & = & \langle \overline{\Obs}; \overline{\Obs} \rangle_{\rm can} 
                \nonumber \\
      & = & (\sigma_{\Obs_i}^2)^{\rm can} 
               \frac{ 2 \tau_{\Obs}^{\rm can} }{N_m}
\end{eqnarray}
but $\sigma_{\hat{\Obs}}^2 \neq (\sigma_{\Obs_i}^2)^{\rm can} 
= \langle \Obs_i ; \Obs_i \rangle$. 
Hence we define an effective autocorrelation time $\tau_{\Obs}^{\rm eff}$
through
\begin{equation}
    \epsilon^2_{\hat{\Obs}} = 
     (\sigma_{\Obs_i}^2)^{\rm can}\frac{ 2 \tau_{\Obs}^{\rm eff} }{N_m}
        = (\epsilon_{\hat{\Obs}}^2)^{\rm can}
             \frac{\tau_{\Obs}^{\rm eff}}{\tau_{\Obs}^{\rm can}},
\end{equation}
i.e.,
\begin{equation}
    \tau_{\Obs}^{\rm eff} = 
       \frac{\sigma^2_{\hat{\Obs}}}{(\sigma_{\Obs_i}^2)^{\rm can}} 
        \tau_{\Obs}.
\end{equation}
For symmetric distributions and {\em odd} observables we have
$\langle \Obs_i w_i \rangle \equiv 0$ and this simplifies to
\begin{equation}
    \epsilon_{\hat{\Obs}}^2 = 
      \frac{ \langle \Obs_i w_i ; \Obs_i w_i \rangle }
           { \langle w_i \rangle^2 } 2 \tau^{\rm int}_{\Obs w; \Obs w},
\end{equation}
such that
\begin{equation}
     \tau_{\Obs} = \tau^{\rm int}_{\Obs w; \Obs w},
\end{equation}
and
\begin{equation}
     \tau_{\Obs}^{\rm eff} = 
          \frac{\sigma^2_{\hat{\Obs}}}{(\sigma_{\Obs_i}^2)^{\rm can}}
            \tau^{\rm int}_{\Obs w; \Obs w}.
\end{equation}
%
%
\newpage
               
%
%
\newpage
{\Large\bf Table Captions}
%
  \vspace{1cm}
  \begin{description}
    \item[\tt\bf Tab. 1:] 
Canonical simulation:
Integrated and exponential autocorrelation times $\tau^{\rm int}$
and $\tau^{\rm exp}$, and flipping time $\tau^{\rm flip}_{m_1}$ using the
standard Metropolis algorithm (M) or the multigrid W-cycle (W),
$\mu^2 = 1.30$.
    \item[\tt\bf Tab. 2:] 
Multicanonical simulation: 
Integrated and exponential autocorrelation times $\tau^{\rm int}$ and
$\tau^{\rm exp}$ using the standard Metropolis algorithm (M)
or the multigrid W-cycle (W).
    \item[\tt\bf Tab. 3:] 
Multicanonical simulation:
Integrated and exponential autocorrelation times
$\tau^{\rm int}_{m_1e^f}$ and $\tau^{\rm exp}_{m_1e^f}$ using
the standard Metropolis (M) or the multigrid W-cycle (W).
Also listed are
the effective multicanonical variance $\sigma_{\hat{\Obs}}^2$
and the canonical variance $(\sigma_{\Obs_i}^2)^{\rm can}$
for $\Obs = m_1$. From these the effective
autocorrelation time $\tau^{\rm eff}_{m_1}$ for the canonical statistical error
estimate can be computed according to eq.(\ref{eq:taueffodd}).
For comparison, in the last column we also list the ``flipping'' time
$\tau^{\rm flip}_{m_1}$ for the diffusion between the peak maxima.
Same values of $n_e$ as in Table \ref{table:mucatau}.
    \item[\tt\bf Tab. 4:] 
Multicanonical simulation: 
Mean values, variances, and covariance, as well as 
integrated autocorrelation times 
$\tau^{\rm int}_{\Obs w; \Obs w}$,
$\tau^{\rm int}_{w; w}$, and
$\tau^{\rm int}_{\Obs w; w}$ for $\Obs = m_2$ and
$w=\exp(f)$ and $\mu^2 = 1.30$.
These values enter the statistical error estimate 
eq.(\ref{eq:fullerror}) for the even observable $m_2$
which allows to compute the squared canonical error estimates
$\epsilon^2$ and the corresponding effective autocorrelation time 
$\tau^{\rm eff}_{m_2}$ defined in eq.(\ref{eq:taueff}).
For comparison the same quantities were also obtained by direct
jackkniving (jack).
    \item[\tt\bf Tab. 5:] 
Interface tension $\sigma_L$ for various lattice sizes 
and $\mu^2 = 1.30$, $1.35$, and $1.40$. The infinite volume
interface tension $\sigma_{\infty}$ was obtained by a fit according to
$\sigma_L = \sigma_{\infty}+a/L$.
 \end{description}
%
\small
\begin{table}[bhp]
\catcode`?=\active \def?{\kern\digitwidth}
\centerline{
\begin{tabular}{|r|r|r||l|l|l||r|l|l|}
\hline
&&
\multicolumn{4}{|c|}{~~~~~~~$\Obs = m_1$}
&\multicolumn{3}{|c|}{$\Obs = m_2$}\\
\hline
\hline
$L$ & 
&
$n_e$ &
???$\tau^{\rm int}_{m_1}$&
???$\tau^{\rm exp}_{m_1}$&
???$\tau^{\rm flip}_{m_1}$&
$n_e$ &
???$\tau^{\rm int}_{m_2}$ &
???$\tau^{\rm exp}_{m_2}$ \\
\hline
 ?4 & M &  3 & ?154.6(3.7) & ?159(11)   & ?200.5(2.4) & 
           3 &   12.366(82) & ?15.07(34) \\
 ?4 & W &  1 & ??15.15(19) & ??15.43(36)& ??14.49(14) & 
           1 &   ?1.4794(87)& ??2.549(82) \\
\hline
 ?8 & M & 20 & ?847(19)    & ?877(54)   & 1007(11)    & 
          20 &   28.19(15)  & ?40.2(1.3) \\
 ?8 & W &  1 & ??40.20(81) & ??39.8(21) & ??48.24(48) & 
           1 &   ?1.841(16) & ??3.70(14)  \\
\hline
 16 & M &150 & 5780(110)   & 5710(320)  & 6575(62)    & 
           2 &   55.4(2.3)  & 118(14) \\
 16 & W &  8 & ?239.6(3.8) & ?248(12)   & ?275.9(2.3) & 
           1 & ?3.475(43) & ??8.06(27)\\
\hline
\end{tabular}
}
\caption[]{%
Canonical simulation:
Integrated and exponential autocorrelation times $\tau^{\rm int}$
and $\tau^{\rm exp}$, and flipping time $\tau^{\rm flip}_{m_1}$ using the
standard Metropolis algorithm (M) or the multigrid W-cycle (W),
$\mu^2 = 1.30$.
}
\label{table:canonical} 
\end{table}
\normalsize 
%
%
%
%
\begin{table}[bhp]
\catcode`?=\active \def?{\kern\digitwidth}
\centerline{
\begin{tabular}{|r|r|r||l|l||l|l|}
\hline
&&&
\multicolumn{2}{|c|}{~~~~~$\Obs = m_1$}
&\multicolumn{2}{|c|}{$\Obs = m_2$}\\
\hline
\hline
$L$ & 
&
$n_e$ &
???$\tau^{\rm int}_{m_1}$&
???$\tau^{\rm exp}_{m_1}$&
???$\tau^{\rm int}_{m_2}$ &
???$\tau^{\rm exp}_{m_2}$ \\
\hline
\multicolumn{7}{|l|}{$\mu^2 = 1.30$}\\
\hline
 ?8 & M &  5 &?204.4(4.0) & ?212(12)    & ?40.73(45)   & ??53.0(1.9) \\
    & W &  1 &??10.88(12) & ??11.30(32) & ??2.542(20)  & ???4.51(12) \\
\hline
 16 & M & 20 &?690(11)    & ?668(23)    & 116.8(1.3)   & ?195.4(6.8)  \\
    & W &  1 &??34.69(76) & ??37.2(2.0) & ??6.224(69)  & ??11.92(42)  \\
\hline
 32 & M & 50 &2984(63)    & 3120(200)   & 390.2(6.7)   & ?899(56)     \\
    & W &  2 &?150.0(4.0) & ?148(11)    & ?20.36(54)   & ??48.5(3.6)  \\
\hline
 64 & M &    &$-$         & $-$         & $-$          & $-$         \\
    & W &  2 &?758(37)    & ?746(62)    & ?78(13)      & ?204(92)     \\
\hline
\multicolumn{7}{|l|}{$\mu^2 = 1.35$}\\
\hline
 ?8 & M &  5 &?209.3(4.0) & ?207.1(9.8) & ?43.92(44)   & ??56.2(1.6) \\
    & W &  1 &??11.48(11) & ??11.42(30) & ??2.870(20)  & ???4.72(13) \\
\hline
 16 & M & 20 &?796(14)    & ?764(31)    & 135.7(1.5)   & ?225.8(8.1) \\
    & W &  1 &??45.26(80) & ??46.9(2.2) & ??8.18(13)   & ??15.23(54) \\
\hline
 32 & M & 50 &4180(130)   & 4590(420)   & 496(13)      & 1220(110)   \\
    & W &  2 &?225.2(7.6) & ?222(18)    & ?28.0(1.2)   & ??70.1(7.6) \\
\hline
 64 & M &    &$-$         & $-$         & $-$          & $-$         \\
    & W &  2 &2130(160)   & 2200(450)   & 128(53)      & ?390(210)    \\
\hline
\multicolumn{7}{|l|}{$\mu^2 = 1.40$}\\
\hline
 ?8 & M &  5 &?240.1(4.0) & ?251(15)    & ?47.93(57)   & ??62.4(2.0) \\
    & W &  1 &??13.11(16) & ??13.15(40) & ??3.326(27)  & ???5.58(16) \\
\hline
 16 & M & 20 &?930(20)    & ?914(49)    & 155.8(2.2)   & ?265(12)    \\
    & W &  1 &??57.4(1.5) & ??61.5(4.2) & ?10.40(19)   & ??19.18(96) \\
\hline
 32 & M & 50 &6050(160)   & 5700(380)   & 641(25)      & 1690(200)   \\
    & W &  2 &?450(19)    & ?460(50)    & ?44.6(2.4)   & ?124(21)    \\
\hline
 64 & M &    &$-$         & $-$         & $-$          & $-$         \\
    & W &  5 &3400(270)   & 3000(630)   & 194(75)      & ?820(780)   \\
\hline
\end{tabular}
}
\caption[]{%
Multicanonical simulation: 
Integrated and exponential autocorrelation times $\tau^{\rm int}$ and
$\tau^{\rm exp}$ using the standard Metropolis algorithm (M)
or the multigrid W-cycle (W).
}
\label{table:mucatau}
\end{table}
%
%
%
%
\small
\begin{table}[bhp]
\centering
\catcode`?=\active \def?{\kern\digitwidth}
\centerline{
\begin{tabular}{|r|r||l|l||l|l||l|l|}
\hline
&&\multicolumn{2}{|c|}{~~~~~~~$\Obs = m_1\exp(f)$}
&&&\multicolumn{2}{|c|}{$\Obs = m_1$}\\
\hline
\hline
$L$ & 
&
???$\tau^{\rm int}_{m_1e^f}$ &
???$\tau^{\rm exp}_{m_1e^f}$ &
???$\sigma^2_{\hat{\Obs}}$ &
???$(\sigma_{\Obs_i}^2)^{\rm can}$ &
???$\tau^{\rm eff}_{m_1}$ &
???$\tau^{\rm flip}_{m_1}$ \\
\hline
\multicolumn{7}{|l|}{$\mu^2 = 1.30$}\\
\hline
 ?8 & M & ?171.1(3.4) & ?209(12)    & 0.9439(14)  & 0.50041(94) & ??322.7(6.1)
& ??463.5(6.4) \\
    & W & ???9.82(11) & ??11.34(33) & 0.94396(98) & 0.50063(71) & ???18.51(20) 
& ???30.82(25) \\
\hline
 16 & M & ?509.8(8.9) & ?655(31)    & 1.0739(27)  & 0.43515(58) & ?1258(21)
& ?1759(24) \\
    & W & ??27.58(59) & ??36.9(2.0) & 1.0661(36)  & 0.43606(82) & ???67.4(1.3)
& ???91.7(1.3) \\
\hline
 32 & M & 1840(40)    & 2880(190)   & 1.3102(80)  & 0.3982(13)  & ?6050(120) 
& ?7780(140) \\
    & W & ??96.6(2.4) & ?146(13)    & 1.3304(95)  & 0.39910(64) & ??321.9(7.6)
& ??428.2(8.9) \\
\hline
 64 & M & $-$         & $-$         & $-$         & $-$         & $-$ &
$-$ \\
    & W & ?374(23)    & ?600(120)   & 1.782(39)   & 0.38692(71) & ?1724(86) & 
?1922(85) \\
\hline
\multicolumn{7}{|l|}{$\mu^2 = 1.35$}\\
\hline
 ?8 & M & ?164.9(3.0) & ?211(11)    & 1.3005(37)  & 0.5824(11)  & ??368.1(6.0)
& ??517.1(7.5) \\
    & W & ???9.925(88)& ??11.47(34) & 1.3013(19)  & 0.58324(64) & ???22.14(20)
& ???35.71(30) \\
\hline
 16 & M & ?521(11)    & ?790(45)    & 1.7065(69)  & 0.54426(71) & ?1635(32)
& ?2088(31) \\
    & W & ??32.02(66) & ??48.3(2.6) & 1.6775(92)  & 0.54649(72) & ???98.3(1.9)
& ??125.1(2.0) \\
\hline
 32 & M & 1821(48)    & 4370(340)   & 2.861(22)   & 0.53264(86) & ?9780(240)
& 11140(240) \\
    & W & ?103.1(4.9) & ?253(32)    & 3.016(39)   & 0.53298(49) & ??584(26)
& ??664(18) \\
\hline
 64 & M & $-$         & $-$         & $-$          & $-$        & $-$ 
& $-$ \\
    & W & ?622(48)    & 2090(400)   & 3.70(12)     & 0.5289(39) & ?4350(320)
& ?4570(310) \\
\hline
\multicolumn{7}{|l|}{$\mu^2 = 1.40$}\\
\hline
 ?8 & M & ?176.4(4.1) & ?250(17)    & 1.6672(53)  & 0.66704(96) & ??440.9(9.7)
& ??581.3(8.9) \\
    & W & ??10.73(14) & ??13.03(45) & 1.6762(43)  & 0.66560(58) & ???27.02(32)
& ???41.66(38) \\
\hline
 16 & M & ?530(12)    & ?940(59)    & 2.458(16)   & 0.64361(60) & ?2017(41)
& ?2451(39) \\
    & W & ??35.47(93) & ??59.8(4.1) & 2.409(20)   & 0.64430(62) & ??132.6(3.0)
& ??158.3(2.9)  \\
\hline
 32 & M & 2215(59)    & 5330(440)   & 3.709(43)   & 0.6357(15)  & 12920(320)
& 14620(360) \\
    & W & ?167.5(7.1) & ?426(56)    & 3.806(56)   & 0.63657(46) & ?1001(40)
& ?1065(35) \\
\hline
 64 & M & $-$         & $-$         & $-$         & $-$        & $-$ &
$-$ \\
    & W & ?778(63)    & 3330(530)   & 6.90(22)    & 0.6275(82) & ?8550(600)
& ?8780(530) \\
\hline
\end{tabular}
}
\caption[]{%
Multicanonical simulation:
Integrated and exponential autocorrelation times
$\tau^{\rm int}_{m_1e^f}$ and $\tau^{\rm exp}_{m_1e^f}$ using
the standard Metropolis (M) or the multigrid W-cycle (W).
Also listed are
the effective multicanonical variance $\sigma_{\hat{\Obs}}^2$
and the canonical variance $(\sigma_{\Obs_i}^2)^{\rm can}$
for $\Obs = m_1$. From these the effective
autocorrelation time $\tau^{\rm eff}_{m_1}$ for the canonical statistical error
estimate can be computed according to eq.(\ref{eq:taueffodd}).
For comparison, in the last column we also list the ``flipping'' time
$\tau^{\rm flip}_{m_1}$ for the diffusion between the peak maxima.
Same values of $n_e$ as in Table \ref{table:mucatau}.
}
 \label{table:taueffodd}
\end{table}
\newpage
\normalsize
%
%
%
%

\begin{table}[bhp]
\catcode`?=\active \def?{\kern\digitwidth}
\centerline{
\begin{tabular}{|r|r||l|l|l|l|}
\hline
$L$ & 
&
??$\langle \Obs_i w_i \rangle$ &
???$\langle w_i \rangle$ &
$\langle \Obs_i w_i; \Obs_i w_i \rangle$ &
??$\langle w_i; w_i \rangle$ \\
\hline
\hline
 ?8 & M & 0.3676(14) & 0.4626(14) & 0.12856(40) & 0.12231(23) \\
    & W & 0.3679(10) & 0.4629(11) & 0.12862(30) & 0.12249(17) \\
\hline
 16 & M & 0.2711(11) & 0.3474(13) & 0.10067(33) & 0.13448(38) \\
    & W & 0.2739(17) & 0.3507(20) & 0.10127(49) & 0.13508(56) \\
\hline
 32 & M & 0.1972(17) & 0.2541(21) & 0.08095(55) & 0.12383(81) \\
    & W & 0.1928(18) & 0.2487(23) & 0.07926(61) & 0.12139(89) \\
\hline
 64 & M & $-$        & $-$        & $-$         & $-$         \\
    & W & 0.1388(35) & 0.1791(45) & 0.0654(14)  & 0.1056(23)  \\
\hline
\hline
$L$ & 
&
$\langle \Obs_i w_i; w_i \rangle$ &
???$\tau^{\rm int}_{\Obs w; \Obs w}$ &
???$\tau^{\rm int}_{w; w}$ &
???$\tau^{\rm int}_{\Obs w; w}$ \\
\hline
\hline
 ?8 & M & 0.12159(29) & ?47.98(48) & ?51.18(47) & ?50.88(49)  \\
    & W & 0.12173(22) & ??3.467(27)& ??3.758(27)& ??3.672(27) \\
\hline
 16 & M & 0.11526(35) & 161.5(1.9) & 170.4(1.9) & 167.2(1.9)  \\
    & W & 0.11585(52) & ??9.711(89)& ?10.248(92) & ?10.041(90) \\
\hline
 32 & M & 0.09986(66) & 644.1(7.2) & 666.9(7.3) & 656.9(7.2)  \\
    & W & 0.09783(74) & ?35.77(61) & ?37.01(62) & ?36.47(62)  \\
\hline
 64 & M & $-$        & $-$        & $-$        & $-$          \\
    & W & 0.0830(18) & 149.1(5.7) & 151.9(5.8) & 150.7(5.8)   \\
\hline
%
%
\hline
$L$ & 
&
$\epsilon^2 \times 10^6$ &
$\tau^{\rm eff}_{m_2}$ &
$\epsilon^2 \times 10^6$ (jack)&
$\tau^{\rm eff}_{m_2}$ (jack) \\
\hline
\hline
 ?8 & M & 0.540(22)  & ??32.0(1.3)  & 0.445  & ?26.4  \\
    & W & 0.2442(55) & ???2.902(64) & 0.216  & ??2.571 \\
\hline
 16 & M & 0.1064(97) & ??84.3(7.7)  & 0.0947 & ?75.0  \\
    & W & 0.176(16)  & ???6.97(62)  & 0.154  & ??6.09  \\
\hline
 32 & M & 0.035(16)  & ?250(110)    & 0.0398 & 283   \\
    & W & 0.056(27)  & ??16.0(7.6)  & 0.0492 & ?14.0  \\
\hline
 64 & M & $-$        & $-$          & $-$    & $-$   \\
    & W & $-0.22$(13)& $-$240(140)  & 0.0295 & ?32.4  \\
\hline
\end{tabular}
}
\caption[]{%
Multicanonical simulation: 
Mean values, variances, and covariance, as well as 
integrated autocorrelation times 
$\tau^{\rm int}_{\Obs w; \Obs w}$,
$\tau^{\rm int}_{w; w}$, and
$\tau^{\rm int}_{\Obs w; w}$ for $\Obs = m_2$ and
$w=\exp(f)$ and $\mu^2 = 1.30$.
These values enter the statistical error estimate 
eq.(\ref{eq:fullerror}) for the even observable $m_2$
which allows to compute the squared canonical error estimates
$\epsilon^2$ and the corresponding effective autocorrelation time 
$\tau^{\rm eff}_{m_2}$ defined in eq.(\ref{eq:taueff}).
For comparison the same quantities were also obtained by direct
jackkniving (jack).
}
\label{table:taueffeven}
\end{table}
%
%
%
%
\begin{table}[bhp]
\catcode`?=\active \def?{\kern\digitwidth}
\centerline{
\begin{tabular}{|r||l|l|l|}
\hline
$L$ &
$\mu^2 = 1.30$ &
$\mu^2 = 1.35$ &
$\mu^2 = 1.40$ \\
\hline
\hline
        8 & 0.14826(58) &  0.19013(52) & 0.23668(79) \\
\hline
       16 & 0.10526(47) &  0.15634(51) & 0.21288(64) \\
\hline
       32 & 0.07095(39) &  0.12690(42) & 0.18964(49) \\
\hline
       64 & 0.05173(37) &  0.11260(50) & 0.17732(61) \\
\hline
\hline
 $\infty$ & 0.03443(47)  & 0.09785(60)  & 0.16577(73) \\
\hline
\end{tabular}
}
\caption[]{%
Interface tension $\sigma_L$ for various lattice sizes 
and $\mu^2 = 1.30$, $1.35$, and $1.40$. The infinite volume
interface tension $\sigma_{\infty}$ was obtained by a fit according to
$\sigma_L = \sigma_{\infty}+a/L$.
}
\label{table:sigma} 
\end{table}
%
\clearpage
{\Large\bf Figure Captions}
%
  \vspace{1cm}
  \begin{description}
    \item[\tt\bf Fig. 1:] 
Interfaces in a typical mixed phase configuration for 
$\mu^2 = 1.40$ and $L = 64$. Values of $\phi_i > 0.5 (<-0.5)$ have been 
depicted black (white), values of $|\phi_i| < 0.5$ have been depicted 
by varying grey shades. For this configuration the magnetization was
$m_1 \approx - 0.2$.
    \item[\tt\bf Fig. 2a-c:]
Flat multicanonical distributions as compared to the 
canonical double-peak histograms of $m_1$ after reweighting for 
$\mu^2 = 1.30$ and different lattice sizes $L= 8, 16$, and $32$.
    \item[\tt\bf Fig. 3a-c:]
Flat multicanonical distributions as compared to the 
canonical double-peak histograms of $m_1$ after reweighting for 
$L=64$ and $\mu^2 = 1.30, 1.35$, and $1.40$. The arrow in Fig. 3c
indicates the value of $m_1$ which was measured for the 
configuration displayed in Fig. 1. 
    \item[\tt\bf Fig. 4a-d:]
Evolution series of $m_1$ for $L=16$ and $\mu^2=1.30$ using the
canonical Metropolis (Fig. 4a), the multicanonical Metropolis
(Fig. 4b), the canonical multigrid W-cycle (Fig. 4c), and the 
multicanonical multigrid W-cycle (Fig. 4d).
In Figs. 4a and 4c we also show the digitalized time series
according to the simple two-state flip model.
The time scales were adjusted so that an evolution over roughly 
$30\tau^{\rm int}_{m_1}$ is displayed in each figure. 

    \item[\tt\bf Fig. 5:]
Autocorrelation time $\tau(k)$ together with a 
three-para\-meter fit according to eq. (\ref{eq:Bfit}) for
$\Obs = m_1 \exp(f)$, $\mu^2 = 1.40$, $L = 64$, and the multicanonical
W-cycle. 
The horizontal line shows the integrated autocorrelation time 
$\tau^{\rm int}_{m_1e^f}=\tau(\infty)$ together with error bounds. 
The inset shows a logarithmic plot of the autocorrelation function 
$A(j)$ together with linear fits
$\ln A(j) = {\rm const}-j/\tau^{\rm exp}_{m_1e^f}$. 
See the text for a detailed explanation.
    \item[\tt\bf Fig. 6:]
Effective autocorrelation times $\tau^{\rm eff}_{m_1}$ and
integrated autocorrelation times $\tau^{\rm int}_{m_1e^f}$ as a function of $L$.
Straight lines are fits according to $\tau = a L^z$.
    \item[\tt\bf Fig. 7a-f:]
Two-dimensional histograms of $m_1$ and $m_2$ simulated for
$L=64$ and $\mu^2 = 1.40$ and reweighted to different values of $\mu^2$.
The multicanonical distributions are shown in Fig. 7a for $\mu^2=1.40$ 
without reweighting, in Fig. 7c after reweighting to $\mu^2=1.375$, and 
in Fig. 7e after reweighting to $\mu^2=1.35$. Figs. 7b, 7d, and 7f
show the respective canonical distributions after additional 
reweighting with the multicanonical reweighting factor $\exp(f(m))$.
    \item[\tt\bf Fig. 8:]
Interface tension $\sigma_L$ as a function of $\mu^2$ for 
$L=8,16,32$, and $64$. The filled circles show the actual simulation data
and the dashed lines were obtained by reweighting.
The values for $\sigma = \sigma_\infty$ were obtained
by an extrapolation according to $\sigma_L = \sigma_{\infty} + a/L$,
and the solid straight line shows a fit 
$\sigma_{\infty} = a ( \mu^2-\mu^2_c )$ with $\mu^2_c = 1.274(3)$.
 \end{description}
\begin{figure}[p]
\vskip 12.5truecm
\includegraphics{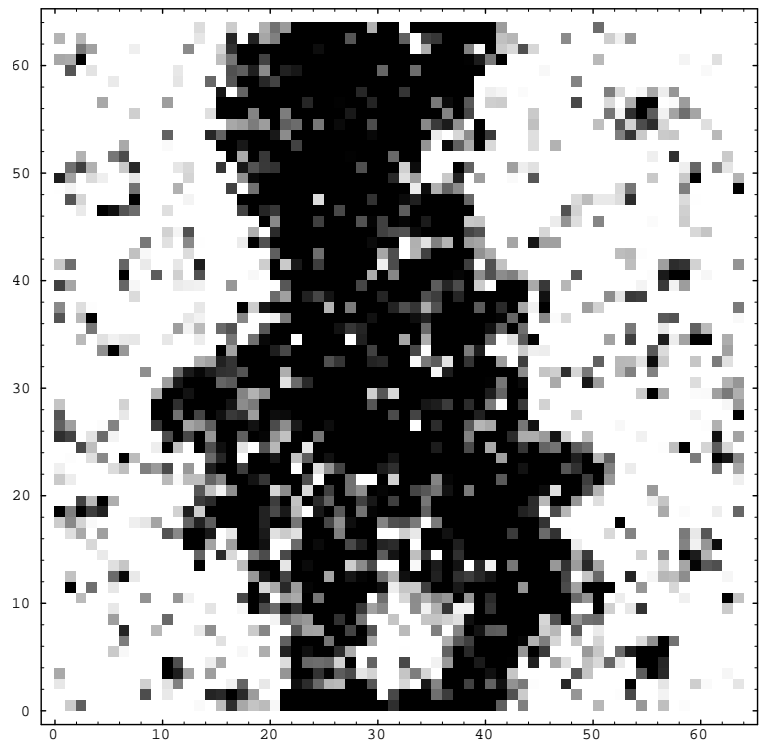}
\caption{%
Interfaces in a typical mixed phase configuration for 
$\mu^2 = 1.40$ and $L = 64$. Values of $\phi_i > 0.5 (<-0.5)$ have been 
depicted black (white), values of $|\phi_i| < 0.5$ have been depicted 
by varying grey shades. For this configuration the magnetization was
$m_1 \approx - 0.2$.
}
\label{fig:1}
\end{figure}
\begin{figure}[p]
\vskip 5.8truecm
\includegraphics{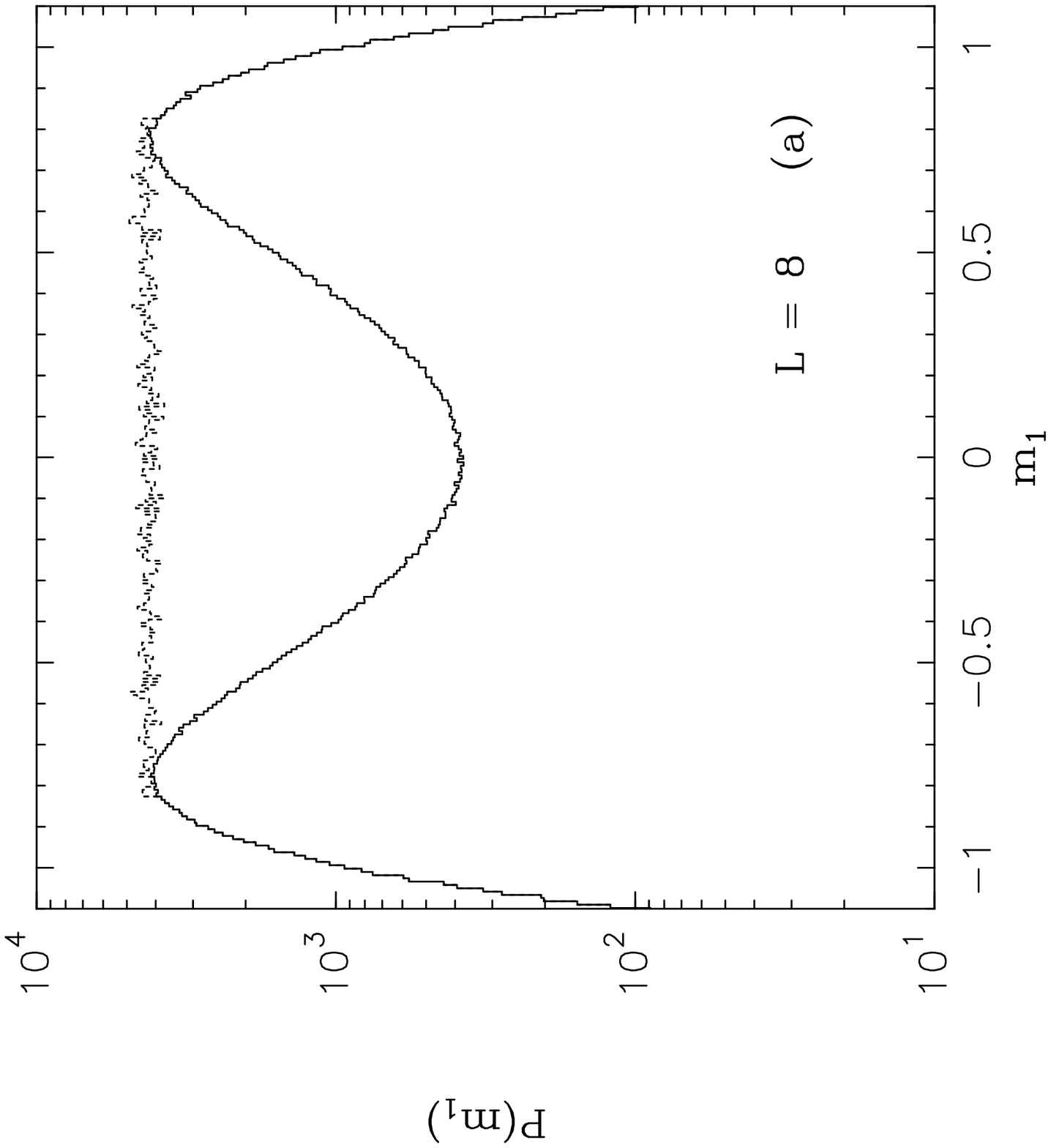}
\vskip 5.8truecm
\includegraphics{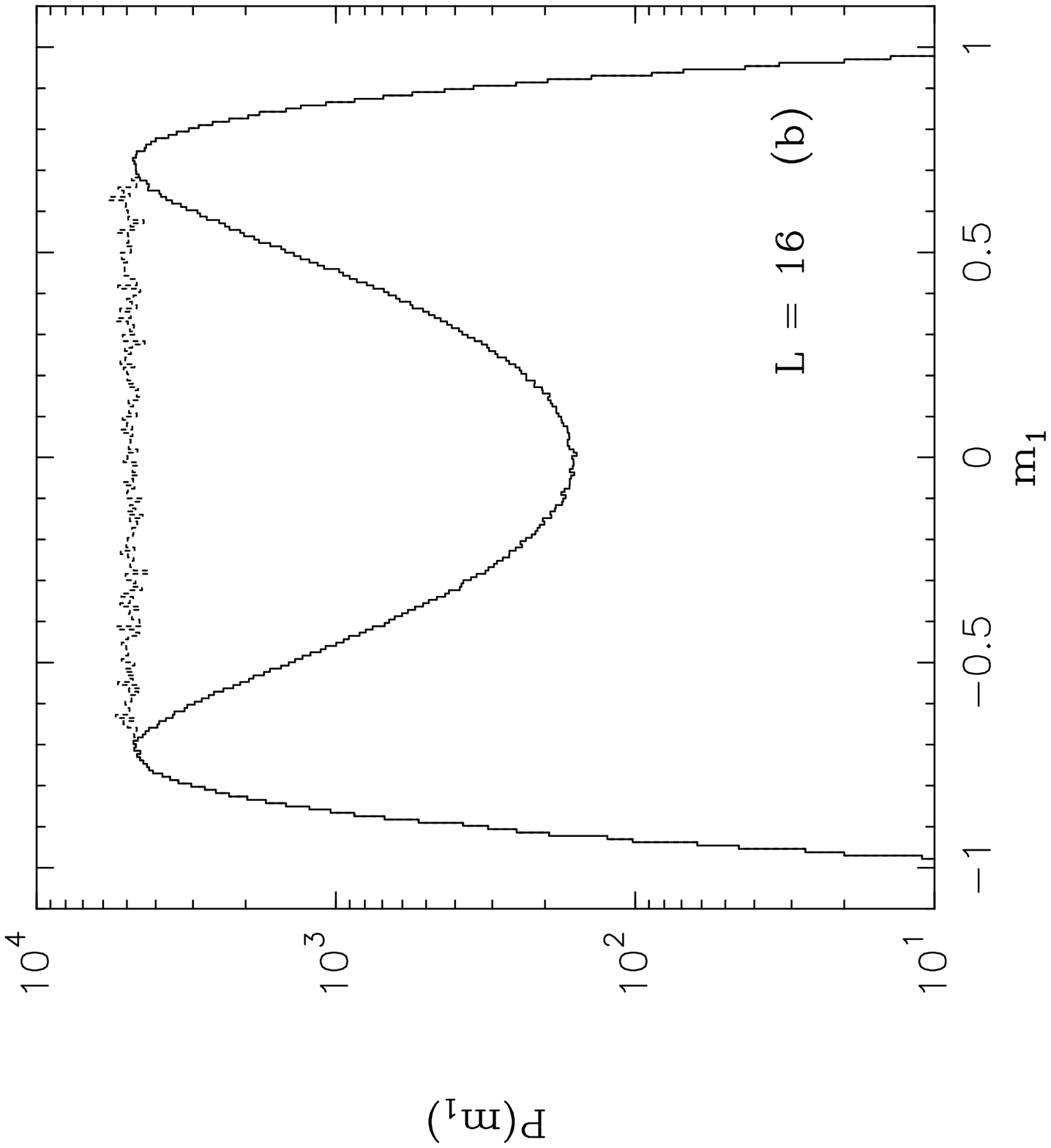}
\vskip 5.8truecm
\includegraphics{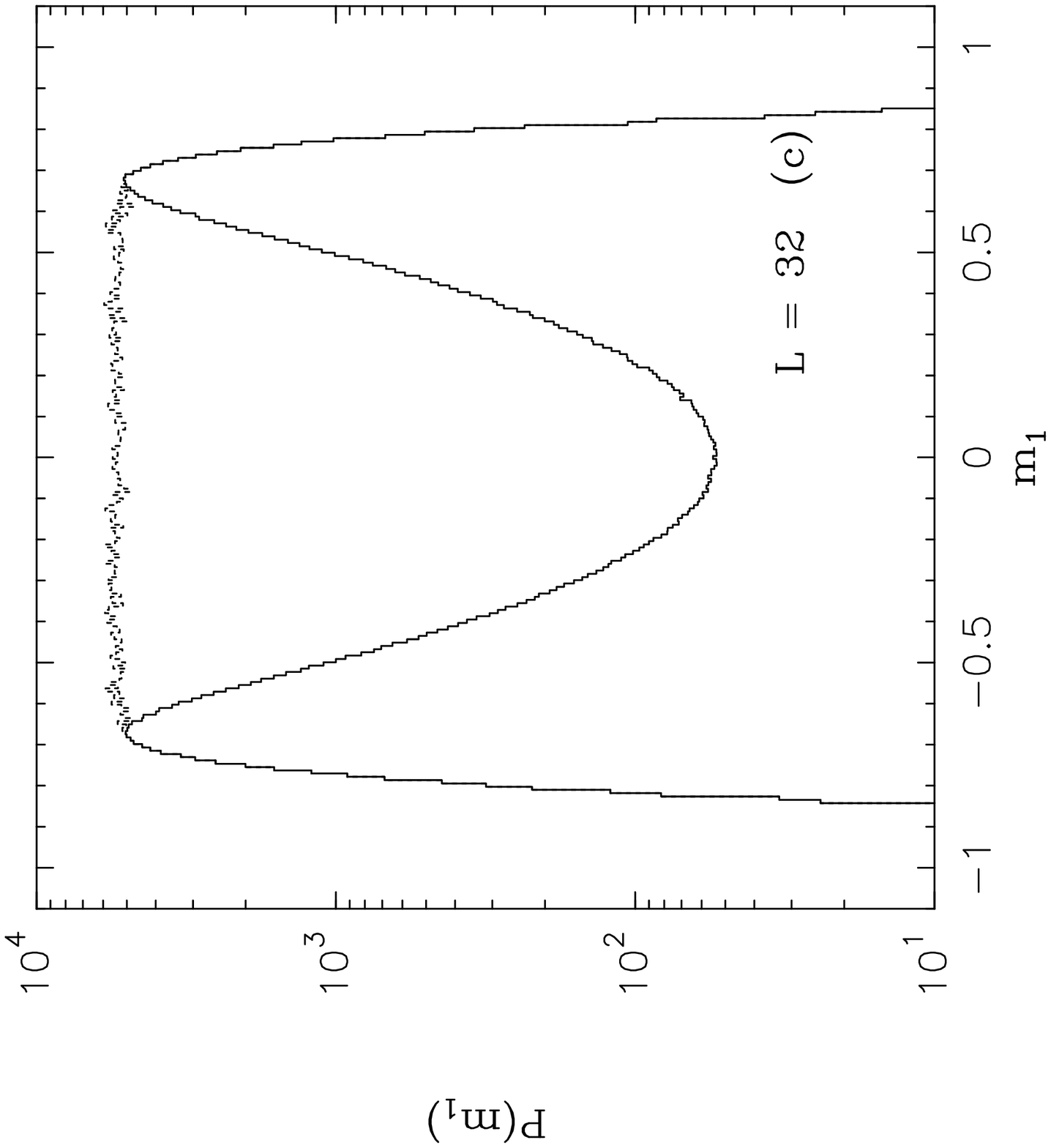}
\caption{%
Flat multicanonical distributions as compared to the 
canonical double-peak histograms of $m_1$ after reweighting for 
$\mu^2 = 1.30$ and different lattice sizes $L= 8, 16$, and $32$.
}
\label{fig:2}
\end{figure}
\begin{figure}[p]
\vskip 5.7truecm
\includegraphics{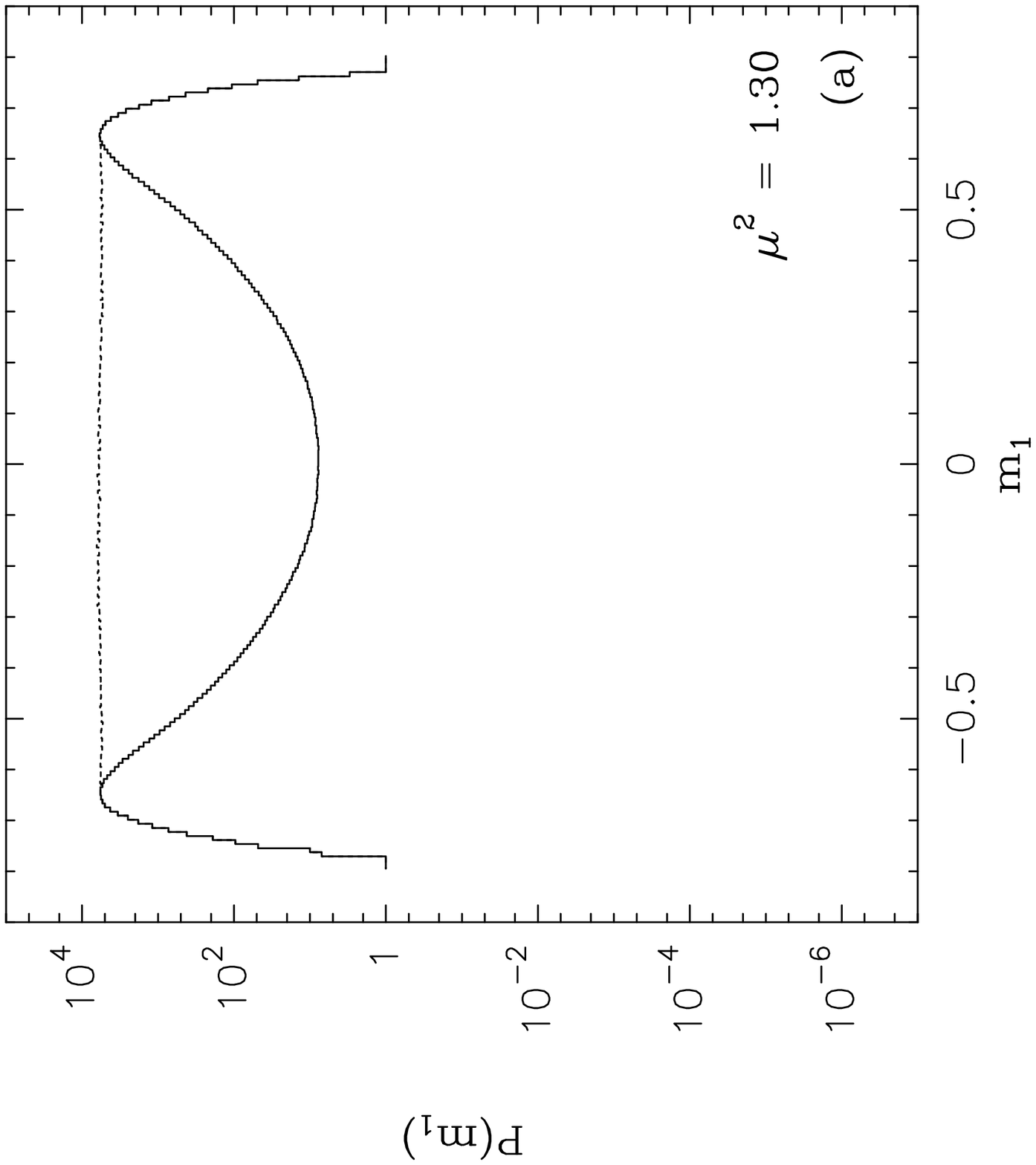}
\vskip 5.7truecm
\includegraphics{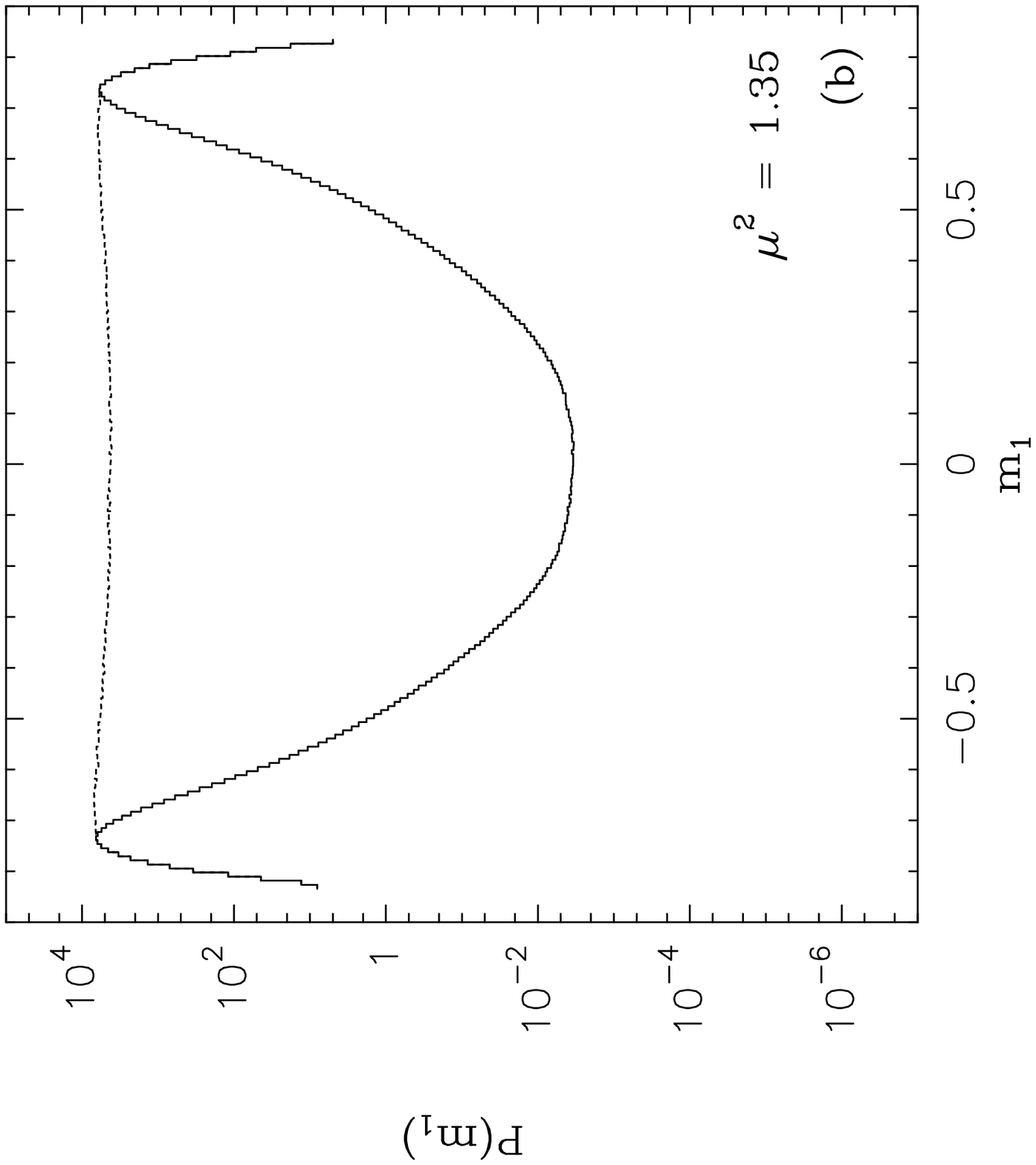}
\vskip 5.7truecm
\includegraphics{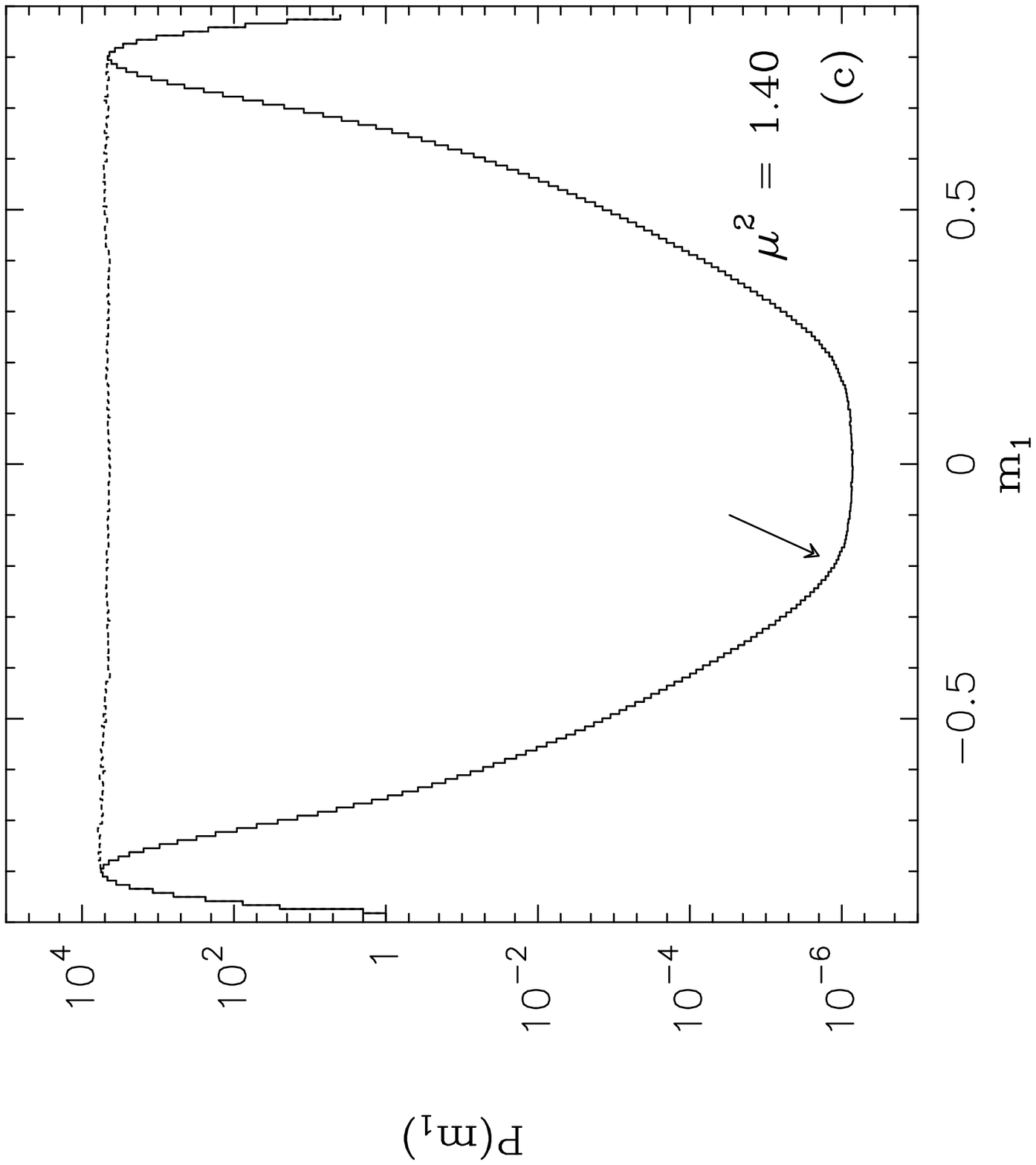}
\caption{%
Flat multicanonical distributions as compared to the 
canonical double-peak histograms of $m_1$ after reweighting for 
$L=64$ and $\mu^2 = 1.30, 1.35$, and $1.40$. The arrow in Fig. 3c
indicates the value of $m_1$ which was measured for the 
configuration displayed in Fig. 1. 
}
\label{fig:3}
\end{figure}
\begin{figure}[p]
\vskip 4.0truecm
\includegraphics{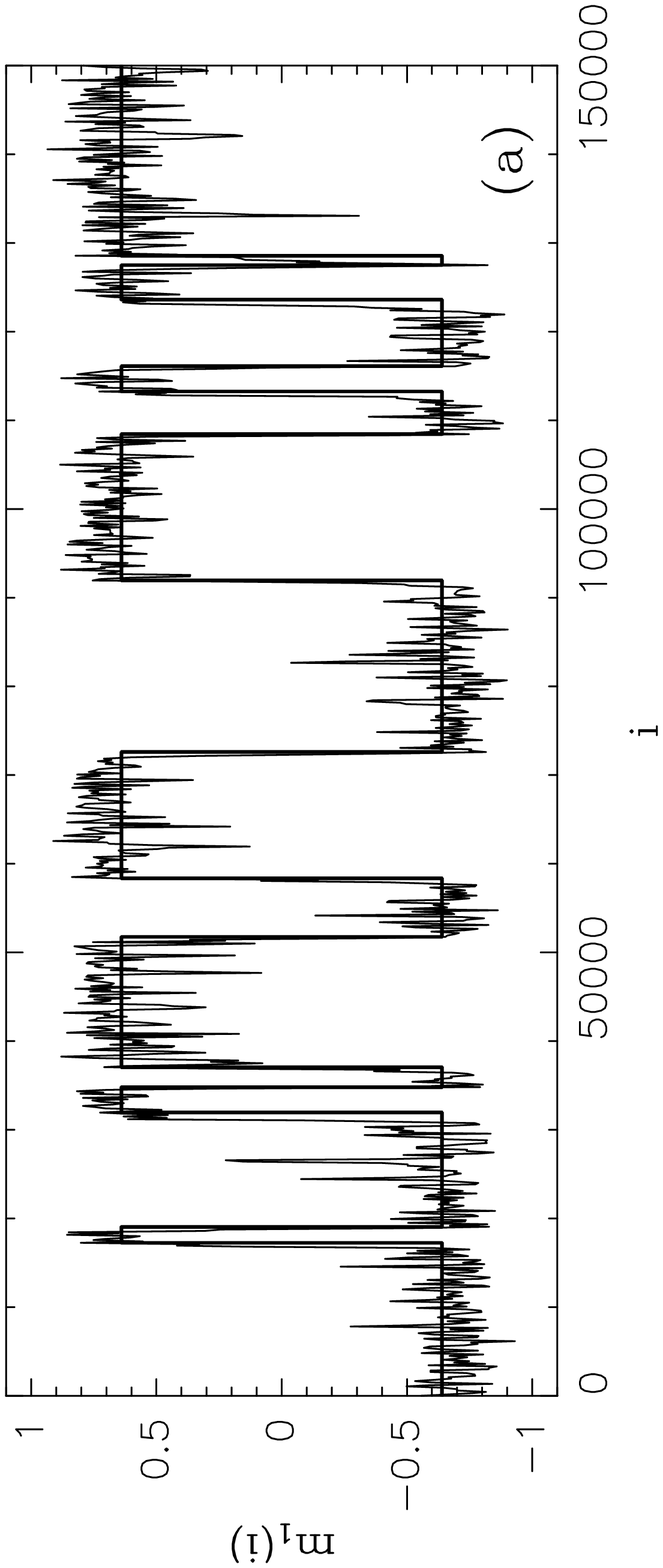}
\vskip 4.0truecm
\includegraphics{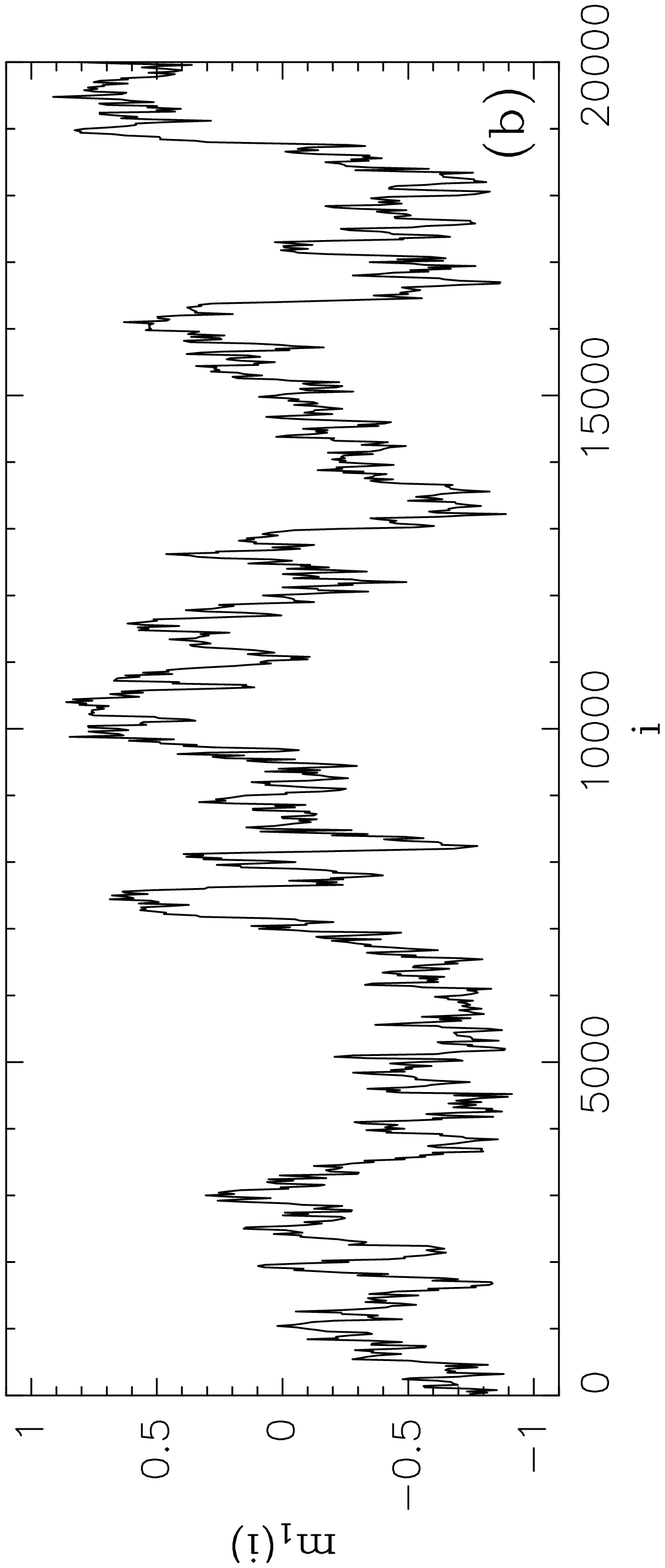}
\vskip 4.0truecm 
\includegraphics{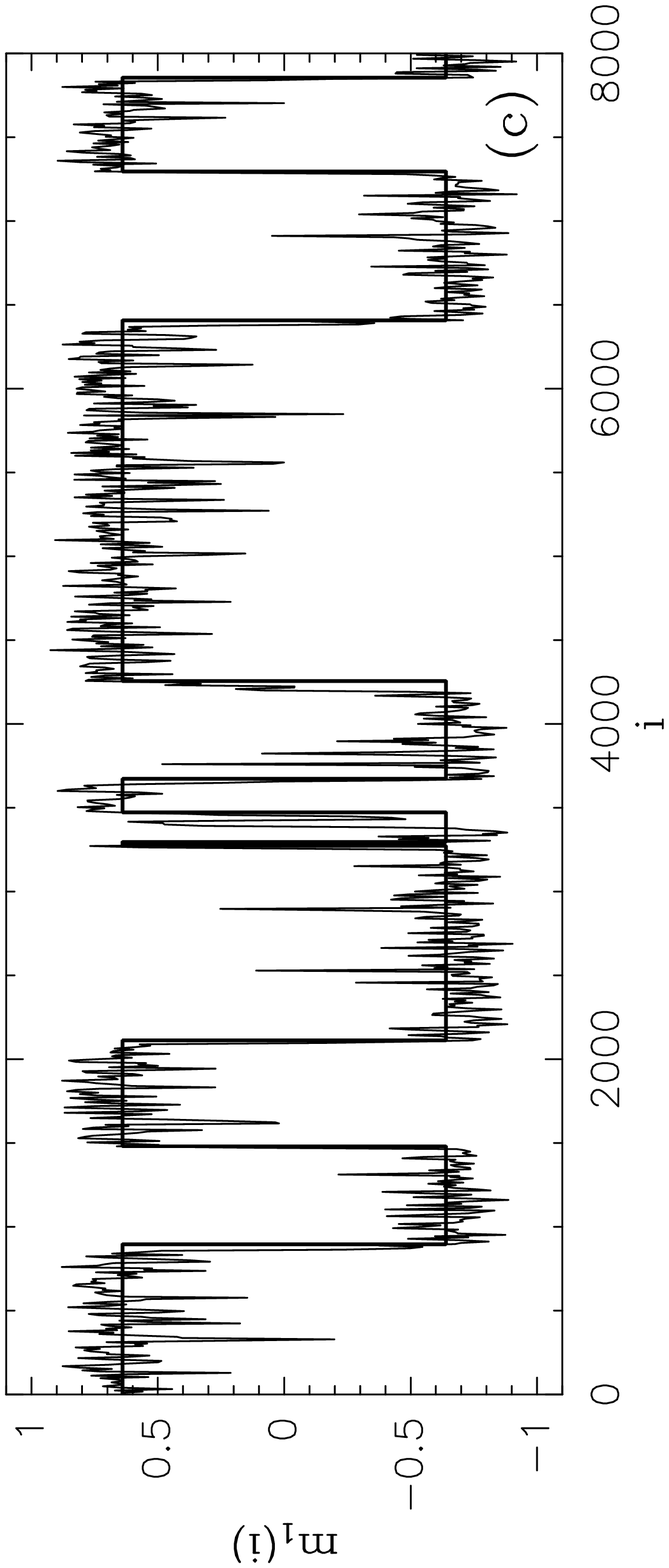}
\vskip 4.0truecm
\includegraphics{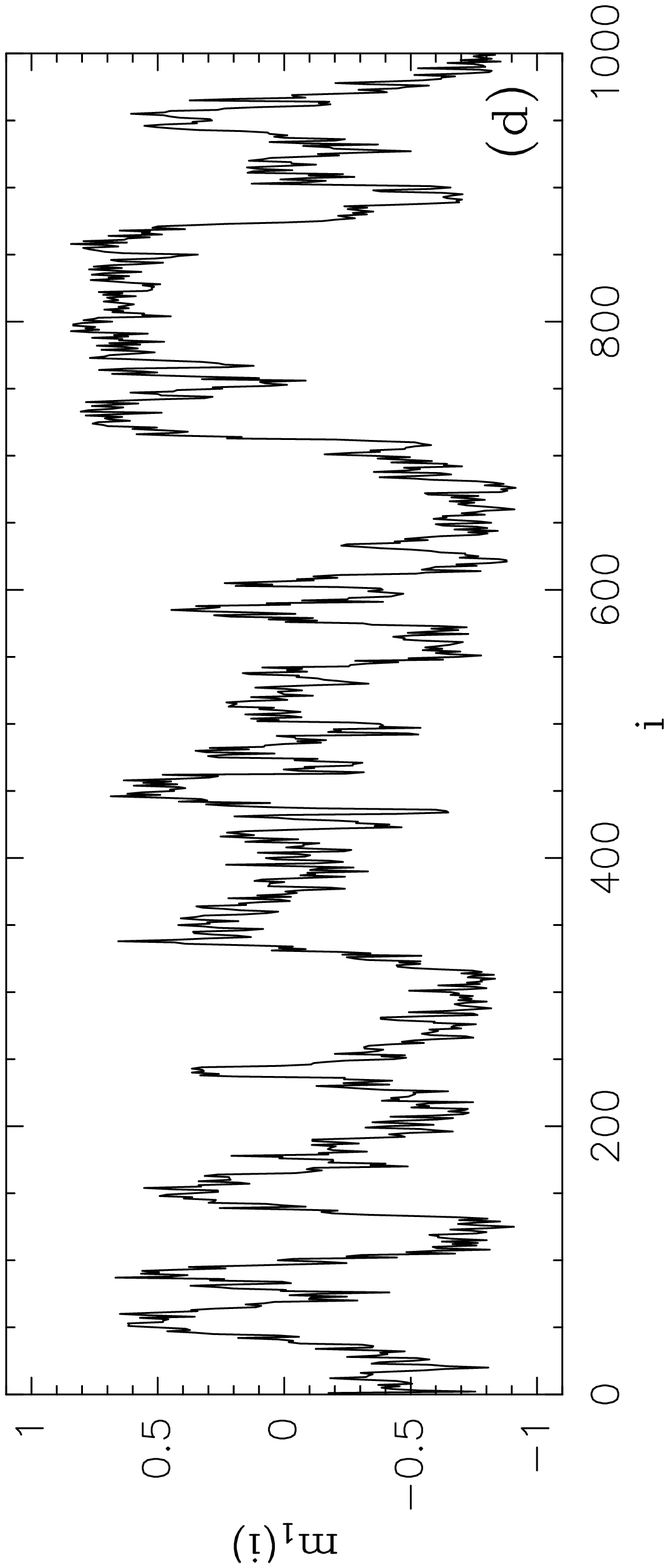}
\caption{%
Evolution series of $m_1$ for $L=16$ and $\mu^2=1.30$ using the
canonical Metropolis (Fig. 4a), the multicanonical Metropolis
(Fig. 4b), the canonical multigrid W-cycle (Fig. 4c), and the 
multicanonical multigrid W-cycle (Fig. 4d).
In Figs. 4a and 4c we also show the digitalized time series
according to the simple two-state flip model.
The time scales were adjusted so that an evolution over roughly 
$30\tau^{\rm int}_{m_1}$ is displayed in each figure. 
}
\label{fig:4}
\end{figure}
\begin{figure}[bh]
\vskip 7.0truecm
\includegraphics{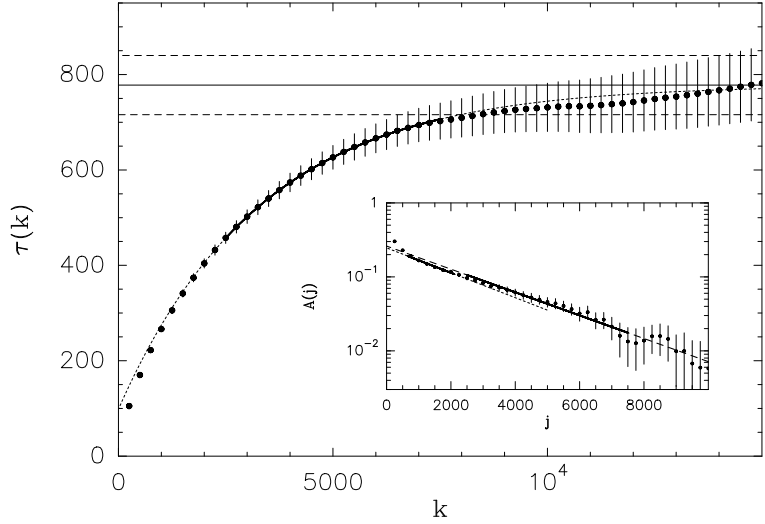}
\caption{
Autocorrelation time $\tau(k)$ together with a 
three-para\-meter fit according to eq. (16) for
$\Obs = m_1 \exp(f)$, $\mu^2 = 1.40$, $L = 64$, and the multicanonical
W-cycle. 
The horizontal line shows the integrated autocorrelation time 
$\tau^{\rm int}_{m_1e^f}=\tau(\infty)$ together with error bounds. 
The inset shows a logarithmic plot of the autocorrelation function 
$A(j)$ together with linear fits
$\ln A(j) = {\rm const}-j/\tau^{\rm exp}_{m_1e^f}$. 
See the text for a detailed explanation.
}
\label{fig:5}
\end{figure}
\begin{figure}[bh]
\vskip 6.5truecm
\includegraphics{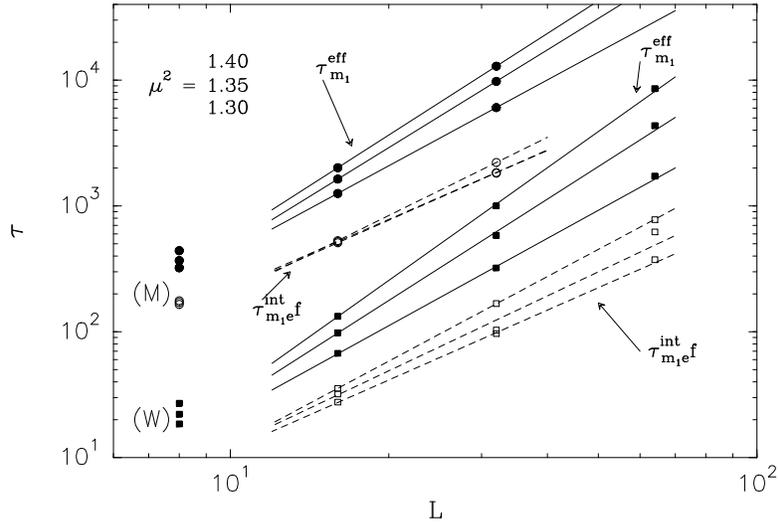}
\caption{%
Effective autocorrelation times $\tau^{\rm eff}_{m_1}$ and
integrated autocorrelation times $\tau^{\rm int}_{m_1e^f}$ as a function of $L$.
Straight lines are fits according to $\tau = a L^z$.
}
\label{fig:6}
\end{figure}
\begin{figure}[p]
\vskip 5.0truecm
\vskip 5.0truecm
\vskip 5.0truecm
\caption{%
Two-dimensional histograms of $m_1$ and $m_2$ simulated for
$L=64$ and $\mu^2 = 1.40$ and reweighted to different values of $\mu^2$.
The multicanonical distributions are shown in Fig. 7a for $\mu^2=1.40$ 
without reweighting, in Fig. 7c after reweighting to $\mu^2=1.375$, and 
in Fig. 7e after reweighting to $\mu^2=1.35$. Figs. 7b, 7d, and 7f
show the respective canonical distributions after additional 
reweighting with the multicanonical reweighting factor $\exp(f(m))$.
}
\label{fig:7}
\end{figure}
\begin{figure}[p]
\vskip 7.0truecm
\includegraphics{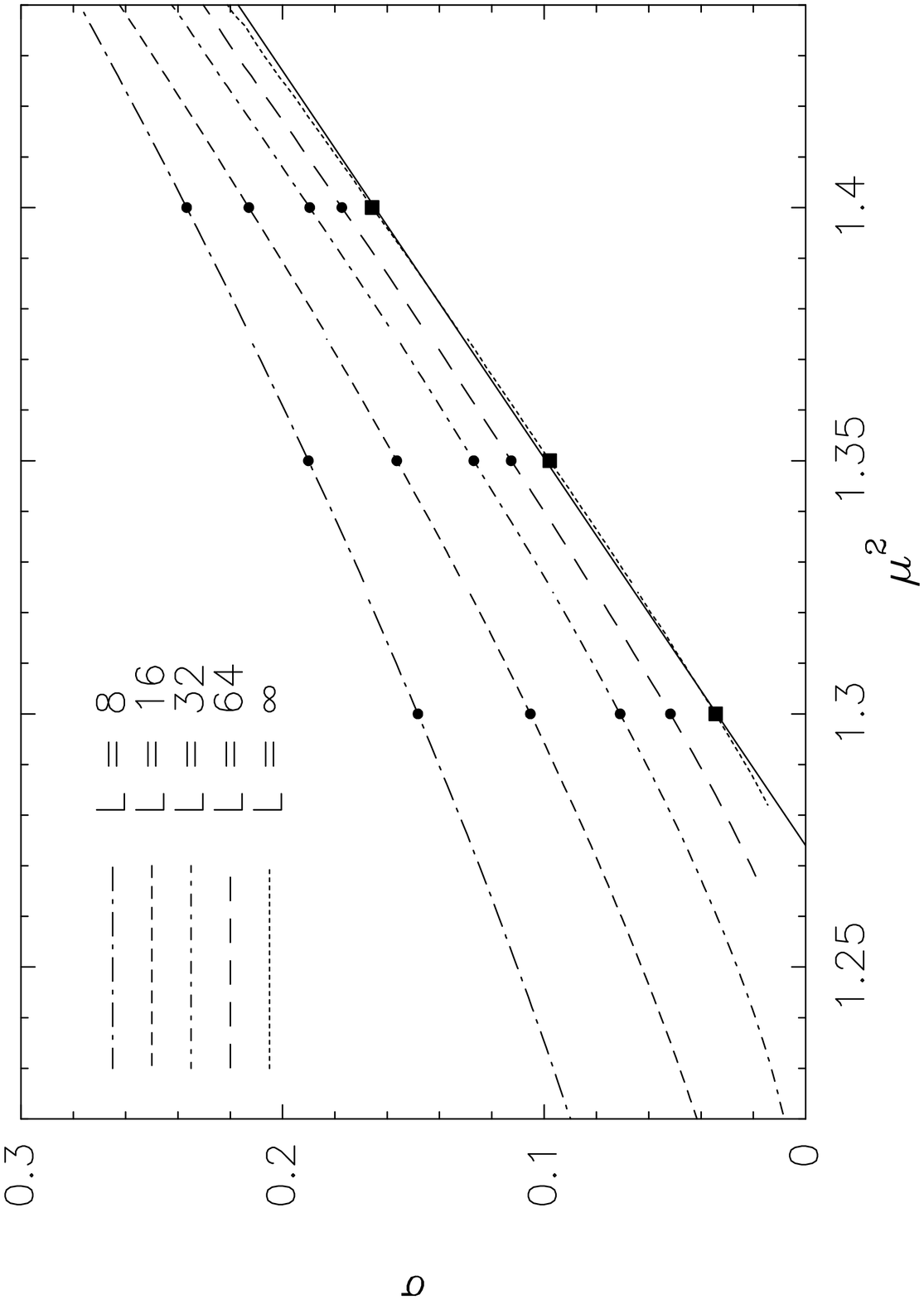}
\caption{%
Interface tension $\sigma_L$ as a function of $\mu^2$ for 
$L=8,16,32$, and $64$. The filled circles show the actual simulation data
and the dashed lines were obtained by reweighting.
The values for $\sigma = \sigma_\infty$ were obtained
by an extrapolation according to $\sigma_L = \sigma_{\infty} + a/L$,
and the solid straight line shows a fit 
$\sigma_{\infty} = a ( \mu^2-\mu^2_c )$ with $\mu^2_c = 1.274(3)$.
}
\label{fig:8}
\end{figure}

\begin{thebibliography}{19}
%
\bibitem{gunton}
%
J.D. Gunton, M.S. Miguel, and P.S. Sahni, in 
{\em Phase Transitions and Critical Phenomena}, Vol. 8, 
edited by C. Domb and J.L. Lebowitz (Academic Press, N.Y., 1983).
%
\bibitem{juelich92}
%
H.J. Herrmann, W. Janke, and F. Karsch (eds.),
        {\em Dynamics of First Order Phase Transitions }
        (World Scientific, Singapore, 1992).
%
\bibitem{muca1}
%
B.A. Berg and T. Neuhaus,
       Phys. Lett. {\bf B267} (1991) 249;
%
%
       Phys. Rev. Lett. {\bf 68} (1992) 9.
%
\bibitem{jbk92}
%
W. Janke, B.A. Berg and M. Katoot,
       Nucl. Phys. {\bf B382} (1992) 649.
%
\bibitem{multimagnetical}
%
B.A. Berg, U. Hansmann, and T. Neuhaus, Phys. Rev. {\bf B47} (1993) 497;
B.A. Berg, U. Hansmann, and T. Neuhaus, Z. Phys. {\bf B90} (1993) 229;
U. Hansmann, B.A. Berg, and T. Neuhaus, in Ref.\cite{juelich92}.
%
\bibitem{muca2a}
%
A. Billoire, T. Neuhaus, and B.A. Berg, 
Nucl. Phys. {\bf B396} (1993) 779;  {\bf B413} (1994) 795.
%
\bibitem{muca2b}
%
B. Grossmann and M.L. Laursen, in Ref.\cite{juelich92}; and
Nucl. Phys. {\bf B408} (1993) 637.
B. Grossmann, M.L. Laursen, T. Trappenberg,
and U.-J. Wiese, Phys. Lett. {\bf B293} (1992) 175.
%
\bibitem{muca3}
%
B.A. Berg, in Ref.\cite{juelich92}.
%
\bibitem{muca4}
%
W. Janke, in Ref.\cite{juelich92}.
%
\bibitem{review}
%
C.F. Baillie, Int. J. Mod. Phys. {\bf C1} (1990) 91;
%
R.H. Swendsen, J.-S. Wang, and A.M. Ferrenberg, {\em New 
Monte Carlo Methods for Improved Efficiency of Computer 
Simulations in Statistical Mechanics\/}, in
{\em The Monte Carlo Method in Condensed Matter Physics\/}, edited 
by K. Binder (Springer, Berlin, 1992) p.~75;
%
A.D. Sokal, 
{\em Bosonic Algorithms}, in {\em Quantum Fields on the Computer},
ed. M. Creutz (World Scientific, Singapore, 1992) p.~211; 
%
A.D. Kennedy, Nucl. Phys. {\bf B} (Proc. Suppl.) {\bf 30} (1993) 96.
%
\bibitem{mgmc1}
%
J. Goodman and A. D. Sokal, Phys. Rev. Lett. {\bf 56} (1986) 1015;
                         Phys. Rev. {\bf D40} (1989) 2035.
%
\bibitem{mgmc2}
%
G. Mack, in {\em Nonperturbative quantum field theory\/}, 
Carg\`ese 1987, ed. G.`t Hooft et al. (Plenum, New York, 1988); 
G. Mack and S. Meyer, Nucl. Phys. {\bf B} (Proc. Suppl.) {\bf 17} 
(1990) 293.
%
\bibitem{mgmc3}
%
D. Kandel, E. Domany, D. Ron, A. Brandt, and E. Loh, Jr., 
        Phys. Rev. Lett. {\bf 60} (1988) 1591; 
D. Kandel, E. Domany, and A. Brandt, 
        Phys. Rev. {\bf B40} (1989) 330.
%
\bibitem{mgmc4}
%
R.G. Edwards, J. Goodman, and A.D. Sokal, Nucl. Phys. {\bf B354} (1991)
289; A. Hulsebos, J. Smit, and J.C. Vink, Nucl. Phys. {\bf B356} (1991)
775; R.G. Edwards, S.J. Ferreira, J. Goodman, and A.D. Sokal, 
Nucl. Phys. {\bf B380} (1992) 621; M.L. Laursen and J.C. Vink, 
Nucl. Phys. {\bf B401} (1993) 745.
%
\bibitem{js93a}
%
W. Janke and T. Sauer,
        Chem. Phys. Lett. {\bf 201} (1993) 499.
%
%
\bibitem{js93b}
%
W. Janke and T. Sauer,
       in {\em Path Integrals from meV to MeV}, ed. H. Grabert et al.
       (World Scientific, Singapore, 1993) p.~17.
%
\bibitem{js93c}
%
W. Janke and T. Sauer,
Phys. Rev. {\bf E49} (1994) 3475. 
%
\bibitem{rummukainen93}
%
K. Rummukainen, Nucl. Phys. {\bf B390} (1993) 621.
%
\bibitem{kw93}
%
W. Kerler and A. Weber, Phys. Rev. {\bf B47} (1993) 11563.
%
\bibitem{mhb86}
%
A. Milchev, D.W. Heermann, and K. Binder,
        J. Stat. Phys. {\bf 44} (1986) 749.
%
\bibitem{bt89}
%
R.C. Brower and P. Tamayo,
        Phys. Rev. Lett. {\bf 62} (1989) 1087.
%
\bibitem{tc90}
%
R. Toral and A. Chakrabarti, Phys. Rev. {\bf B42} (1990) 2445.
%
\bibitem{mf92}
%
B. Mehlig and B.M. Forrest, Z. Phys. {\bf89} (1992) 89;
B. Mehlig, A.L.C. Ferreira, and D.W. Heermann, 
        Phys. Lett. {\bf B291} (1992) 151.
%
\bibitem{binder81}
%
K. Binder, Z. Phys. {\bf B43} (1981) 119, 
           Phys. Rev. {\bf A25} (1982) 1699.
%
\bibitem{ms88}
%
N. Madras and A.D. Sokal,
        J. Stat. Phys. {\bf 50}, 109 (1988).
%
\bibitem{umbrella}
%
G.M. Torrie and J.P. Valleau, {\em Chem. Phys. Lett.} {\bf 28} (1974) 
578; {\em J. Comp. Phys.} {\bf 23} (1977) 187; 
I.S. Graham and J.P. Valleau, {\em J. Phys. Chem.} {\bf 94} (1990) 7894;
J.P. Valleau, {\em J. Comp. Phys.} {\bf 96} (1991) 193.
%
\bibitem{hac85}
%
W. Hackbusch, {\em Multi-Grid Methods and Applications\/} 
(Springer, Berlin, 1985).
%
\bibitem{cor88}
%
S.F. McCormick (ed.), 
{\em Multigrid Methods. Theory, Applications, and Supercomputing\/} 
(Dekker, New York, 1988).
%
\bibitem{sokal93}
%
A.D. Sokal, New York University preprint NYU-TH-93/07/02.
%
\bibitem{neuhaus94}
%
T. Neuhaus,
Nucl. Phys. {\bf B} (Proc. Suppl.) {\bf 34} (1994) 667.
%
\bibitem{gp93}
%
M. Grabenstein and K. Pinn, J. Stat. Phys. {\bf 71} (1993) 607.
%
\bibitem{jackknife}
%
R.G. Miller, Biometrika {\bf 61} (1974) 1; B. Efron,
{\em The Jackknife, the Bootstrap and other Resampling Plans\/}
(SIAM, Philadelphia, PA, 1982).
%
\bibitem{flip}
%
A. Billoire, R. Lacaze, A. Morel, S. Gupta, A. Irb\"ack and B. Peterson,
Nucl. Phys. {\bf B358} (1991) 231; W. Janke, unpublished notes.
%
\bibitem{histogramming}
%
A.M. Ferrenberg and R.H. Swendsen, Phys. Rev. Lett. {\bf 61} (1988) 2635;
{\em ibid.} {\bf 63} (1989) 1195; and {\em Erratum, ibid.}
{\bf 63} (1989) 1658. 
%
%
\bibitem{wiese92}
%
U.-J. Wiese, Bern preprint BUTP-92/37.
%
\end{thebibliography}
\end{document}